%% file: novelty_journal_paper.tex
\documentclass[twoside,11pt]{article}
\usepackage{jair, theapa, rawfonts}

\usepackage{amssymb}
\usepackage{array}
\usepackage{color, colortbl}

\newtheorem{assumption}{Assumption}
\usepackage{multirow} 

\usepackage[pdftex]{graphicx}

\usepackage[cmex10]{amsmath}

\usepackage{url}

\jairheading{?}{2013}{?-?}{?}{?}
\ShortHeadings{Using Temporal IDF for Efficient Novelty Detection in Text Streams}
{Karkali, Rousseau, Ntoulas \& Vazirgiannis}
\firstpageno{1}

\begin{document}

\title{Using temporal IDF for efficient novelty detection in text streams}

\author{\name Margarita Karkali \email karkalimar@aueb.gr\\
	Department of Informatics, Athens University of Economics and Business, Greece
        \AND
        \name Fran\c cois Rousseau \email rousseau@lix.polytechnique.fr\\
        Computer Science Laboratory (LIX), Ecole Polytechnique, France
        \AND
        \name Alexandros Ntoulas \email antoulas@di.uoa.gr\\
	University of Athens, Greece\\
        Zynga, USA
        \AND
        \name Michalis Vazirgiannis \email mvazirg@aueb.gr\\
	Department of Informatics, Athens University of Economics and Business, Greece\\
        Computer Science Laboratory (LIX), Ecole Polytechnique, France
}

\maketitle

\begin{abstract}

Novelty detection in text streams is a challenging task that emerges in quite a few different scenarios, ranging from email thread filtering to RSS news feed recommendation on a smartphone. An efficient novelty detection algorithm can save the user a great deal of time and resources when browsing through relevant yet usually previously-seen content. Most of the recent research on detection of novel documents in text streams has been building upon either geometric distances or distributional similarities, with the former typically performing better but being much slower due to the need of comparing an incoming document with all the previously-seen ones. In this paper, we propose a new approach to novelty detection in text streams. We describe a resource-aware mechanism that is able to handle massive text streams such as the ones present today thanks to the burst of social media and the emergence of the Web as the main source of information. We capitalize on the historical \textit{Inverse Document Frequency} (IDF) that was known for capturing well \textit{term specificity} and we show that it can be used successfully at the document level as a measure of \textit{document novelty}. This enables us to avoid similarity comparisons with previous documents in the text stream, thus scaling better and leading to faster execution times. Moreover, as the collection of documents evolves over time, we use a temporal variant of IDF not only to maintain an efficient representation of what has already been seen but also to decay the document frequencies as the time goes by. We evaluate the performance of the proposed approach on a real-world news articles dataset created for this task. 
We examine an exhaustive number of variants of the model and compare them to several commonly used baselines that rely on geometric distances. The results show that the proposed method outperforms all of the baselines while managing to operate efficiently in terms of time complexity and memory usage, which are of great importance in a mobile setting scenario.
\end{abstract}

\input{1_introduction.tex}

\input{2_related.tex}

\input{3_method.tex}

\input{4_dataset.tex}

\input{5_experiments.tex}

\input{6_results.tex}

\input{6.1_extended_model_evaluation.tex}

\input{7_conclusions.tex}

\section*{Acknowledgments}

The research of M. Karkali has been co-financed by the EU (ESF) and Greek national funds through the Operational Program "Education and Lifelong Learning" of the NSRF - Heracleitus II. The research of F. Rousseau and M. Vazirgiannis was partially financed by the French DIGITEO grant LEVETONE.
The research of A. Ntoulas was partially supported by PIRG06-GA-2009-256603.

\bibliographystyle{theapa}
\bibliography{bibliography}

\end{document}

%% file: 1_introduction.tex
\section{Introduction}


A great deal of information consumption these days happens in the form of push notifications: a user
specifies a general topic or stream that he is interested in watching or following and a specific service
sends updates to his email, desktop or smartphone. 
In certain cases, the user may be interested in following {\em all} the stories coming from a specific
source. In others, for example with sources like Twitter, Facebook or certain news sites that allow posting of
variants of a given story, the user might be interested in having a way of specifying that he is interested 
only in stories that he is not aware of or, in other words, only in stories that are {\em novel}.

This problem emerges in a variety of different settings, from email thread filtering to RSS news feed recommendation on a smartphone
and is commonly called {\em first story detection}\footnote{There are additional names in the literature for 
this same problem, i.e. novelty detection, novelty mining, new event detection, topic initiator detection or
new event detection.} (FSD).
A good novelty detection algorithm can potentially save a lot of time to the user (by hiding previously seen
stories and not just posts) but also bandwidth, battery and storage especially in the mobile setting scenario.
Moreover, by proposing a service that can deliver fresh and novel content, this service might increase its user engagement and user retention.
In 2002, a novelty track was introduced in TREC~\cite{TREC} where sentence-level novelty detection 
had to be combined with relevance in order to retrieve novel sentences from a set of relevant ones.

At a high level, previous research on novelty detection focuses on the definition of a similarity (or distance) metric that is
used to compare each new incoming document (e.g. a post or a tweet) to a set of previously seen documents.
If the similarity of the new incoming document is below a threshold (defined differently in each work) then the document
is considered novel and therefore some relevant action (e.g. email to the user) is taken on the document.
The similarity functions used in the literature range in effectiveness and complexity
from simple word counts to cosine similarity through online clustering and one-class classification \cite{UMass,filtering,part1,sentence,hypermedia}.

In prior work, cosine similarity has been reported to work better than most of the previously
proposed approaches~\cite{UMass,sentence,filtering} and was shown to outperform 
even complex language model-based approaches in most cases. 
The documents were represented as bag-of-word vectors with additional
TF$\times$IDF term weighting applied on them.

Although previous approaches have been shown to work well in most cases, they have two shortcomings.
On one hand, the {\em document-to-document approaches} such as the cosine similarity ones~\cite{UMass} tend to be computationally expensive as we need to compare the new incoming document with all the previously seen documents in order to determine its novelty.
If the user wishes to have a reasonably large collection of documents to compare to, this approach can prove very costly for a system supporting 
millions of users or, in the case of a mobile setting, may drain the phone's battery faster.
On the other hand, the {\em document-to-summary approaches} such as the online clustering 
or one-class classification~\cite{TREC,part1}, where we compare the document to a summary (e.g. the centroid of a cluster),
are faster and more appropriate for a mobile setting,
but they were shown to be less effective than the document-to-document approaches~\cite{UMass,TREC}.

To this end, we propose a document-to-summary technique that is both efficient computationally
and effective in performing novelty detection.
Our main idea is to maintain a summary of the collection of previously seen documents
that is based on the specificity of each term. We capture the specificity of each term
through its \textit{Inverse Document Frequency} (IDF) and, for a given incoming document, we then show how to compute
its overall specificity through the definition of a novelty score. Since our approach is document-to-summary based, we do not compare to all the previous documents and thus we can compute the novelty score faster. 
Moreover, because the topics in a news stream may shift over time, we complement our method with a time decay
technique that aims at giving recent documents more weight than older ones when performing novelty detection.
We show in our experimental evaluation that our approach is similar in
performance to the document-to-summary approaches and in certain cases
it beats them by a wide margin.

\vspace{0.1in}

The main contributions of this paper are: 
\begin{itemize}
\item A new document scoring function for novelty detection based on IDF that captures the difference between a new document's vocabulary and the collection's vocabulary seen so far, synonymous with novelty.
\item An extensive experimental evaluation of our proposed method and the
commonly used baselines. Our results indicate that our method outperforms previous ones in both
execution time and precision in identifying novel texts.
\item An annotated corpus that can be used as a benchmark for novelty detection in text streams extracted from a recent real-world news stream.\footnote{The dataset is publicly available at: \url{http://www.db-net.aueb.gr/GoogleNewsDataset/}.}
\end{itemize}

The remainder of the paper is organized as follows. 
Section \ref{sec:related} provides a review of related work.
In Section \ref{sec:method},  we discuss the baseline techniques for novelty detection, we introduce the notation we use throughout the paper
and we introduce our new scoring function for novelty detection. 
In Section \ref{sec:GND} we describe the dataset we constructed for the evaluation of the proposed method.
Section \ref{sec:experiments} describes the experimental setting and the development of 
the dataset used. In Section \ref{sec:results} and \ref{sec:ext_mod_eval}, we present the experimental results and 
the effectiveness of the general and the extended model proposed respectively. Finally, Section \ref{sec:conclusion} closes the work with our conclusions.

%% file: 2_related.tex
\section{Related Work}
\label{sec:related}

In this section, we present the various tasks and approaches that novelty detection encompasses and we review the standard datasets used for evaluation in related work.

\subsection{Various Tasks and Approaches to Novelty Detection}

Novelty detection is usually described as a task in signal processing.  
A survey on methods for novelty detection has been published on Signal Processing Journal by Markou and Singh. 
The survey is separated in two parts: statistical approaches \cite{part1} and neural networks \cite{part2}. 
Novelty detection is a challenging task, with many models that perform well on different data. In this survey,
novelty detection in textual data was reported to be a variant of traditional text classification, and it was mentioned as
an alternative terminology of Topic Detection and Tracking (TDT). 

In the Topic Detection and Tracking (TDT) field, many papers are dealing with the problem of First Story Detection (FSD) that is synonymous with novelty detection in news streams. 
This corresponds to the task of online identification of the earliest report for each event as soon as that report arrives in the sequence of documents. 
In TDT-3 competition \cite{TDT}, which included a FSD task, Allan \textit{et al.} presented a simple 1-NN approach, also known as UMass \cite{UMass} that is reported to perform at least as well as the other participants. 
The UMass approach is constantly used as a baseline in relevant literature.
An interesting report from the FSD task in the context of TDT was also published by Allan \textit{et al.} \citeyear{Allan_Hard_2000}, concluding that FSD based on tracking approaches bounds its performance. In our approach we do not model tracking and thus such limitations do not apply.

An interesting work by Yang \textit{et al.} \citeyear{topic} uses topic clustering, Named Entities (NE) and topic specific stopword removal for the task of novelty detection on news. 
Novelty detection at the document level was also used in adaptive filtering \cite{filtering}, where the document streams are user profiles based on which new documents must be identified as redundant with regard to these profiles. The measures tested were separated between geometric distance and language model measures. The results show that the simple approach of maximum cosine distance, introduced by Allan \textit{et al.} \citeyear{UMass}, work as well as complex language model measures and a mixture model proposed by the authors. An interesting review of novelty detection techniques on adaptive hypermedia systems was presented by Lin and Brustilovisky \citeyear{hypermedia}, reporting the difficulty to tackle the problem using traditional methods. A recent work by Verheij \textit{et al.} \citeyear{Verheij_2012} presents a comparison study of different novelty detection methods evaluated on news article from Yahoo! News Archive where language model-based methods perform better than cosine similarity-based ones. 

In addition to the TDT competition, novelty detection was also present in TREC. In TREC 2002-2004 the novelty track was introduced \cite{trec2002,trec2003,trec2004}. Novelty detection was then examined at the sentence level and the general goal of the track was to highlight sentences that contain both relevant and novel information in a short, topical document stream. A paper by Sobboroff and Harman \citeyear{TREC} reported the significant problem in evaluating such tasks, by highlighting problems in the construction of a groundtruth dataset.

Based on TREC novelty track, a significant amount of work was published on novelty detection at the sentence level \cite{sentence,sl1,sl2,sl3,sl4}. Allan \textit{et al.} \citeyear{sentence} evaluated seven measures for novelty detection separating them in word count measures and language model measures. The results showed that the simple approach of maximum cosine similarity between a sentence and a number of previously seen ones works as well as complex language model measures that are based on smoothing and mixture models. The Meiji University experiments in TREC 2003 \cite{Meiji_2003} proposed a linear combination of the maximum cosine similarity measure with a metric that aggregates the \textit{TF$\times$IDF} scores of the terms in a sentence. This metric is similar to the one presented here, but it is tested for sentence-level novelty detection which is a different task from the one we tackle in the current work. 

Lately the interest in novelty detection and mainly in FSD has been focused on reducing the computation time since FSD is an online task, and the prevalent 1-NN approach relies on exhaustive document-to-document similarity computation. Petrovic \textit{et al.} \citeyear{twitter} approximate 1-NN with LSH (Locality Sensitive Hashing). Zhang \textit{et al.} \citeyear{Zhang_NED_2007} also aims at improving the efficiency of novelty detection systems introducing a news indexing-tree. Luo \textit{et al.} \citeyear{Luo_TED_2007} presents a framework for online new event detection used in a real application. The approach used is the 1-NN approach and the framework focuses on improving system efficiency by reducing the number of saved documents using indices, parallel processing, etc. Our method also manages to increase the efficiency of novelty detection by avoiding the exhaustive comparisons present in 1-NN approach as we describe in the next section.

\subsection{Benchmark Datasets for Novelty Detection}
\label{sec:relatedData}

Novelty detection in text streams is usually evaluated in news applications, as this is the most common form of text streams and the task of finding novel news articles makes perfect sense. 
Most of the work on novelty and first story detection use the TDT datasets for evaluation \cite{TDT}. The TDT benchmark 
collection is  sparsely labeled: the most recent collection from TDT (TDT5) includes 278,109 English news articles but only 100 topics and around 4,500 annotated documents. Another benchmark dataset, mainly for sentence-level novelty detection is the TREC novelty track dataset. This dataset is not adequate for the purpose of this paper as it contains novelty judgments per sentence and not per document. 
Tsai \textit{et al.} \citeyear{Tsai_weighting_2011} used the TREC dataset for document-level novelty scoring using the number of novel sentences per document but we believe that such assumptions are vague and cannot lead to safe conclusions.
Other works \cite{topic,filtering} use available news article collections and apply sampling and manual labeling using known events in a specific time span. For example, from a collection of 261,209 articles, a sample of 538 documents from 36 categories is selected and annotated \cite{topic}.
Details for these datasets are available also in a paper from Tsai \citeyear{Tsai_Review_2010}.

All the above efforts have the common feature that the evaluation datasets are always manually annotated - thus there is always the issue of human subjective judgment that introduces a niche of uncertainty. In addition, in most cases, there is almost no guarantee that the very first story on each annotated event will be included in the annotated documents.

%% file: 3_method.tex
\section{Novelty Scoring Methods}
\label{sec:method}

We consider a system that monitors a stream of documents. New documents arrive at the system at different times. Each document bears a timestamp that corresponds to the time of creation. We assume that documents reach the system ordered by their creating time. Each document $d^t$, with a timestamp $t$, is represented using a \textit{bag-of-word} approach, as $<(q_1,w_1^C),(q_2,w_2^C),...,(q_{|d^t|},w_{|d^t|}^C)>$, where $q_i$ is the $i^{th}$ unique term in document $d^t$ and $w_i^C$ is the corresponding weight computed with regard to a corpus $C$.

\begin{figure}[t]
\centering
\includegraphics[height=3cm]{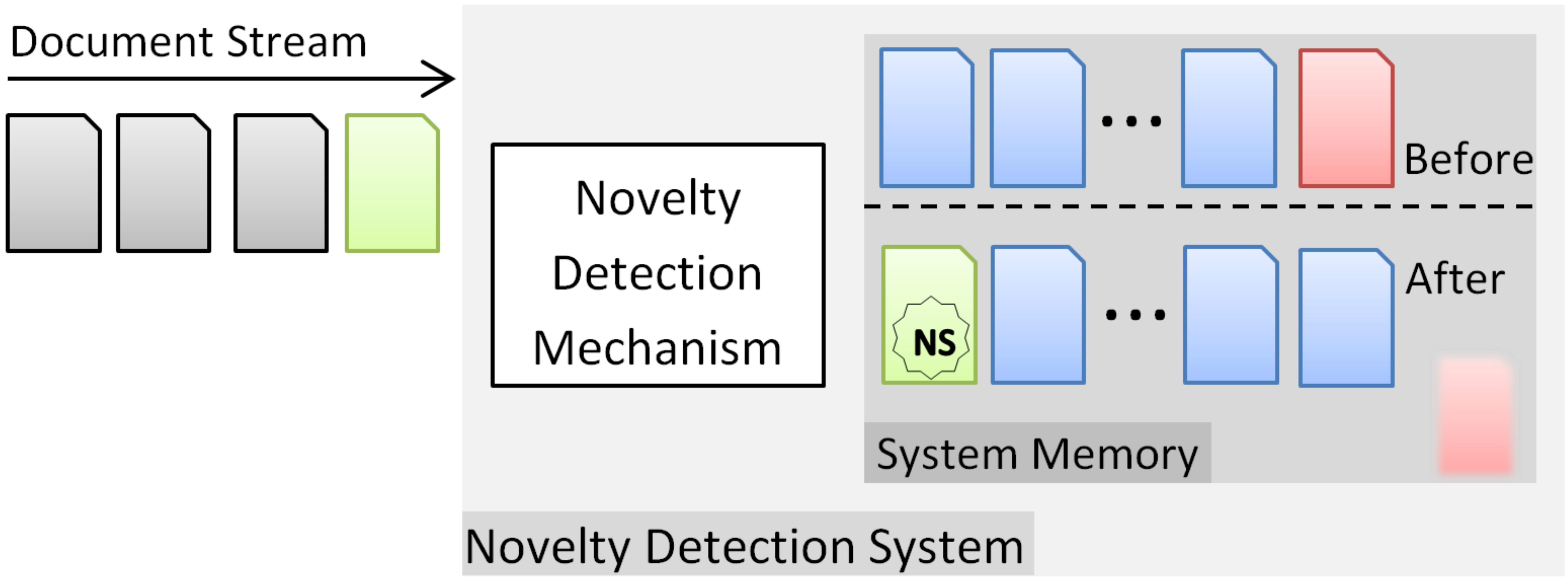}
\caption{The process of Novelty Detection in a text stream.}
\label{fig:model}
\end{figure}

When a new document $d^t$ arrives in the system, the previous $N$ 
 ones are already stored and indexed. 
We use the terms \textit{memory}, \textit{corpus} and \textit{collection} for this set of previously seen documents interchangeably in the rest of the paper. Assuming a corpus $C$, for each new document $d^t$, a novelty score $NS(d^t, C)$ is computed and indicates the novelty of this document with regard to the corpus. When this score is computed, $d^t$ is stored in memory and the oldest document is flushed. This process is illustrated in Figure \ref{fig:model}.

We define the Novelty Detection (ND) problem as the characterization of an incoming document as novel with respect to a predefined window in the past. In the described context we declare novel a document $d^t$ when the corresponding novelty score $NS(d^t, C)$ is higher than a given threshold $\theta$.

\subsection{Baseline Methods}
We present in the next sub-sections the different baselines used in the related work and that we will compare our novel method with; on one hand the \textit{document-to-document approaches} based on either a vector space or a language model and on the other hand the \textit{document-to-summary approaches}.

\subsubsection{Document-to-Document using Vector Space}
\label{cos_sim}

There is a wealth of  geometric similarity measures for novelty detection in document streams, such as the Manhattan and cosine similarities. Cosine similarity is proven to work better in similar tasks and it is frequently used as a baseline in novelty detection. For such measures, in most cases, the document representation used is the \textit{bag-of-word} approach. In order to decide for a document's novelty, its similarity to all documents in memory has to be computed. The dominant method is to calculate the maximum from these similarities and assign the result to the new document as a novelty score. We also introduce a second baseline in which instead of the maximum similarity, the mean similarity is considered to compute the novelty score.

As for the the \textit{Max Cosine Similarity} baseline also referred to as  1-NN approach, it was introduced by Allan \textit{et al.} at TDT3 \cite{UMass} and is also known as the UMass. 
This method is used as a baseline in many papers \cite{twitter,Zhang_NED_2007,hypermedia,filtering} and it is considered the traditional method for document novelty detection.
The  intuition of this metric is that if a new document is very similar to any another in the corpus, the information it contains was seen before and thus the document cannot be considered as novel.  Thus a document with low maximum similarity to any of the documents in memory is considered as novel.  

When the  \textit{Mean Cosine Similarity} is used then a document is marked as novel if its mean similarity to the documents in the corpus is below a threshold.  

Assuming the cosine similarity among two documents  $d$ and $d'$ is as follows: 
\begin{equation}
  CS(d,d') = \frac{\sum_{k=1}^{m}w_k(d)w_k(d')}{\sqrt{\sum_{k=1}^{m}w_k(d)^2\sum_{k=1}^{m}w_k(d')^2}}
\end{equation}
where the $CS(d,d')$ is the cosine similarity between documents $d$ and $d'$, $w_k(d)$ is the weight/importance of the term k in document d and $m$ is the number of common terms among  the two documents, then the  respective similarity formulas for the aforementioned metrics are as follows: 
\begin{equation}
  MaxCS(d^t, C) = \max_{1\leq{i}\leq{|C|}} CS(d^t,d_i)
\end{equation}
\begin{equation}
  MeanCS(d^t, C) = \frac{\sum_{i=1}^{|C|} CS(d^t,d_i)}{|C|}
\end{equation}
Both approaches are simple to implement but their computational complexity depends on the length of the corpus used. In the worst case, where all the terms in $d^t$ occur in all the documents in memory, the complexity is $O(|d^t| \times |C|)$.

\subsubsection{Document-to-Document using Language Models}

A common method to measure the similarity between two documents to use language models. 
A recent comparison study by Verheij \textit{et al.} \citeyear{Verheij_2012}, where a number of methods were used for novelty detection, reports that the best performing method was a document-to-document distance based on language models.  As mentioned in Section \ref{sec:related}, LM-based methods were previously tested with no significant results. 

We used minimum Kullback-Leibler (KL) divergence as a baseline approach for language models. We implemented the method as described in the study from Verheij \textit{et al.}.
Thus, assuming the KL divergence of a document  $d$ given a document $d'$ is as follows: 
\begin{equation}
  KL(\Theta_{d}, \Theta_{d'}) = \sum_{q \in d}{\Theta_{d}(q)\log{\frac{\Theta_{d}(q)}{\Theta_{d'}(q)}}} 
\end{equation}
where $\Theta_{d}$ is the unigram language model on document d and $\Theta_{d}(q)$ is the probability of term q in document d, then the respective novelty scoring formula is as follows: 
\begin{equation}
  MinKL(d^t, C) = \min_{1\leq{i}\leq{|C|}} KL(\Theta_{d^t},\Theta_{d_i})
\end{equation}

In order to avoid the problem of zero probabilities we use linear interpolation smoothing, smoothing document weights against the corpus.
\begin{equation}
  \Theta_{d^t}(q) = \lambda*\Theta_{d^t}(q) + (1-\lambda)*\Theta_{d^1...d^(t-1)}(q)
\end{equation}
where $\lambda$ is the smoothing parameter having a value in [0,1] and $\Theta_{d^1...d^{t-1}}$ is the probability of term q in the corpus. For our experiments $\lambda$ was set to 0.9 following Verheij \textit{et al.}'s recommendations.

Note that this approach has the same computational complexity as the previously described ones since it is also a document-to-document approach.

\subsubsection{Document-to-Summary using Vector Space}

A way to avoid the computationally expensive comparisons of the document with regard to the previously seen ones is to maintain a summary of the past documents and compare the incoming one to this summary only. 

To have a complete set of baselines for the evaluation of our method, we also include a document-to-summary approach based on vector space for the document representation and cosine similarity for the novelty metric. The corpus summary is  defined based on \cite{Verheij_2012} as the concatenation of all the documents in corpus:
\begin{equation}
  D_C = \bigcup_{d\in C}d
\end{equation}
Then the novelty scoring formula for the document-to-summary baseline can be define as follows:
\begin{equation}
  AggCS(d^t, C) = CS(d^t,D_C)
\end{equation}

\subsection{Inverse Document Frequency for Novelty}
\label{sec:NS}

In this subsection, we discuss the \textit{Inverse Document Frequency} (IDF) usually used for \textit{term specificity} and we show how we can use it at the document level to propose a novel document scoring function for novelty detection.

\subsubsection{Design Principles}
In this paper, we introduce a novelty score that \textit{does not use any similarity or distance measure}. We compute a score for the incoming document and we compare it to a pre-defined threshold. This novelty score can be considered as a way to compare a document to a corpus, which is the essence of a novelty detection task.

To do so, we capitalize on the \textit{Inverse Document Frequency} (IDF) introduced by Sparck \textit{et al.} \citeyear{sparck_jones_statistical_1972}. IDF is a heuristic measure of term specificity and is a function of term use. More generally, by aggregating all the IDF of the terms of a document, IDF can be seen as a function of the vocabulary use at the document level. Hence, our idea to use it as an estimator of novelty -- a novel document being more likely to use a different vocabulary than the ones in the previous documents. IDF was initially defined as follows:
\begin{equation}
	idf(q, C) = \log{\frac{N}{df_q }}
\end{equation}
where $q$ is the considered term, $C$ the collection, $df_q$ the document frequency of the term $q$ across $C$ and $N$ the size of $C$, i.e. the number of documents.

There exists a slightly different definition known as \textit{probabilistic} IDF used in particular in BM25 \cite{robertson_okapi_1994} where the IDF is interpreted in a probabilistic way as the odds of the term appearing if the document is irrelevant to a given information need and defined as follows:
\begin{equation}
	idf_{probabilistic}(q, C) = \log{\frac{N - df_q}{df_q }}
\end{equation}
Note that this IDF definition can yield to negative values if the term q appears in more than half of the documents as discussed by Robertson and Walker \citeyear{robertson_relevance_1997}. For ad-hoc information retrieval, it has been claimed that this violates a set of formal constraints that any scoring function should meet \cite{fang_formal_2004} but for novelty detection, this property could be of importance as we want to penalize the use of terms appearing in previously seen documents. We will test both versions in our experiments. Actually, we used in practice smoothed variants to handle extreme cases where the document frequency could be null or equal to the size of the collection since the collection is pretty small (memory of the last 100 news for example) and thus subject to sparseness in terms of vocabulary. For the standard IDF, we used add-half Laplace smoothing and for the probabilistic IDF we used the one in BM25 (formulas s and b in Table \ref{tab:smart_notations}, fourth column).

Consequently, if we consider a set of timestamped documents as a corpus, part of a stream, and a new document arriving from that stream, we are interested in its novelty and we can compute a novelty score with regard to the previously seen documents. In a way, the document is novel if its terms are also novel -- i.e. previously unseen. This implies that the terms of a novel document have a generally high specificity and therefore high IDF values with regard to the corpus $C$. The corpus of previously seen documents is of fixed size $N$, reasonable though to correspond to the memory needed from the application.

\begin{table*}
\centering
\def\arraystretch{1.2}
\resizebox{\textwidth}{!}{%
\begin{tabular}{| l | c || l | >{\centering}m{2cm} || l | c |}
\hline
Notation					& Term frequency			& Notation						& Inverse Document Frequency			& Notation				& Normalization\\ \hline

b (boolean)				&
$\left\{
\begin{array}{l}
  1 \quad \mbox{if } tf > 0\\
  0 \quad \mbox{otherwise}
\end{array}
\right.$
												& t (idf)						& $\log{\frac{N}{df}}$						& n (none)				& $1$\\

\multirow{2}{*}{n (natural)}		& \multirow{2}{*}{$tf$	}		& \multirow{2}{*}{s (smoothed idf)}	 & \multirow{2}{*}{$\log{\frac{N + 1}{df + 0.5}}$}	& \multirow{2}{*}{u (\# unique terms)}	
																																	 & \multirow{2}{*}{$|d|$}\\
						&						&							&									& 					&\\
\multirow{2}{*}{l (logarithm)}	& \multirow{2}{*}{$1 + \log{tf}$}	& \multirow{2}{*}{p (prob. idf)}		& \multirow{2}{*}{$\log{\frac{N - df}{df}}$}		& d ($L^1$ norm)		& $\sum{tf}$\\
						&						&							&									& c ($L^2$ norm)		& $\sqrt{\sum{tf^2}}$\\
\multirow{2}{*}{k (BM25)}		& \multirow{2}{*}{$\frac{(k_1 + 1) \times tf}{k_1 \times (1 - b + b \times \frac{dl}{avdl}) + tf}$}
												& \multirow{2}{*}{b (BM25)	}		& \multirow{2}{*}{$\log{\frac{N - df + 0.5}{df + 0.5}}$}
																												& \multirow{2}{*}{p (pivot)}	& \multirow{2}{*}{$1 - b + b \times dl / avdl$}\\
						&						&							&									&					&\\
\hline
\end{tabular}}
\caption{\label{tab:smart_notations}Extended SMART notations that include BM25 components.}
\label{tab:smart}

\end{table*}

\subsubsection{Novelty Score Definition and Properties}
It seems then natural to define our novelty score as a TF$\times$IDF weighting model since we are relying on a \textit{bag-of-word} representation and a \textit{vector space} model. The task here is more of \textit{filtering} than ad-hoc IR, hence the TF component may not need to be concave. We explored indeed a great variety of combinations for TF and IDF that we will present following the SMART notations: the historical ones defined in \cite{singhal_length_1995} and additional ones that include BM25 components. In general, the novelty score of a new document $d$ for a collection $C$ can be defined as follows:
\begin{equation}
	NS(d, C) = \frac{1}{norm(d)}\sum_{q \in d}{tf(q, d) \times idf(q, C)} \label{eq:novelty_score}
\end{equation}
where $tf$, $idf$ and $norm$ can be any of the functions presented in Table \ref{tab:smart_notations}, ranging from a standalone IDF ($bsn$) to a BM25 score ($kbn$) using the SMART triplet notation.

Note that because of the way BM25 is designed, the length normalization is already included in the TF component ($k\_\_$) for a slop parameter $b$ greater than 0. Therefore, BM25 is denoted by $kbn$. We also extended the SMART notations to include a normalization by the number of unique terms ($\_\_u$) rather than the length of the document typically because it makes more sense when considering a boolean TF ($b\_\_$). Indeed, this model makes no difference between a term occurring once or several times, thus the normalization should not in order to be consistent.

The aggregation (through the sum operation) of the term scores to obtain a document score reduces the impact of synonymy which is a common problem when using \textit{bag-of-word} representation and \textit{vector space} model. Indeed, a document that would have terms synonymous with the ones in the other documents would probably be detected as novel since its terms have high IDF values. Nevertheless, it is very unlikely that all its terms are synonymous and overall, its score should not be as high as the one of a novel document. In practice, the number of \textit{false positives} with this method is far less than with state-of-the-art techniques, which validates this fair assumption.

Unlike the approach described in \ref{cos_sim}, this measure is not related to the size $N$ of the corpus used. Its complexity is $O(|d|)$. In addition, no document vector needs to be retrieved (and a fortiori stored in an inverted index except for $d$) for the computation of $NS$. The index is only used after the score has been assigned in order to decrease the \textit{document frequency} of the terms occurring in the oldest document (the one being flushed). Thus, the response time of the system is not affected.

\subsection{Selecting which Corpus to Use}

The baseline methods that rely on term weighting for document representation needs a corpus on which IDF will be computed. This corpus can either be a static collection of news articles or an evolving corpus. The latter contains the recent articles in a sliding window (SW), defined either on the publication date or on the number of articles.

To avoid introducing additional parameters, most of the recent work \cite{twitter} on online novelty detection equates two different corpora: the one used to compute IDF and the collection of articles with which the new article will be compared to. 
Even though these two concepts are not identical, we adopt this assumption and leave to future work the separate examination of these two corpora.

The proposed method, presented in section \ref{sec:NS}, is meaningful only when the corpus describes the recent past, as this is the property that offers IDF the power to indicate novel information. 
Thus, we  focus on examining the potential of novelty detection methods when the corpus used in an evolving one. 

The length of the SW is a parameter that must be taken under examination. 
Here we can give some intuition about it, but each stream may have special characteristics that can be identified only after a thorough experimentation. 
The choice of SW is basically related to the time necessary to forget an event with no more updates. 
Remember that, unlike Topic Tracking problem, we are not interested in updates on an event previously reported, regardless of the extend of the update. 
Thus, we need to remember as deep in the past as needed in order to keep track of events that may still get an update. 

Based on this thought we make a simple yet intuitive assumption:
\begin{assumption}\label{ass:1}
Given an event $E$ and the temporally ordered set of all its reports in press until now $R = \{d_{t_1}, d_{t_2}, ..., d_{t_n}\}$, where $t_i<t_{i+1}$ for each $0<i<n$, then the probability $p(d_{t_{n+1}})$ of document $d_{t_{n+1}}$ to occur in the news stream at a time $t_{n+1}>t_n$ is inversely proportional to the quantity $\delta=t_{n+1}-t_n$. 
\end{assumption}

The best size for the sliding windows corresponds to the tradeoff between forgetting an event too early (and report any next update as novel -- false alarm error) and remembering events too old (and discard novel articles as updates because they may "look like" an old one -- miss detection error).

\subsection{Term Weighting for Evolving Corpora}
\label{sec:tdf}

Selecting the right window size is a difficult task that implies a strict restriction: discarding all articles exiting the window but remembering as equal all the other regardless of their position in the stream. 

We want to incorporate the assumption previously introduced in a method that alleviate this restriction in order to use it alongside with the proposed Novelty Scoring method ($NS$).
As $NS$ works on term level instead of document level, we capitalize on the $tDF$ function we introduced in a previous paper \cite{tdf}. $tDF$ replaces the traditional $DF$ in cases of evolving corpora, weighting terms not only based on the number of occurrences but also on their distribution in the corpus. Here, we employ a more generic form of $tDF$ than the one introduced in \cite{tdf}. For a given term $q$ and a time $t$, we define:
\begin{equation}
	tDF(q, t) = tDF(q, t_{old}) \times f(t - t_{old}) + 1
\end{equation}
where $t_{old} \in [0,t-1]$ is the time of term $q$'s last occurrence and $f$ a given function.
When term $q$ first shows up, $t_{old}$ is undefined. In this case, we set $tDF(q, t)=1$. With $tDF$, terms that are frequent in the recent past will receive higher $tDF$.

Here we need to stress the importance of the decay function $f$ which controls the $tDF$ according to a term's occurrences in the past. At a high level, when a term first occurs, it is indexed with an initial weight of 1. This weight is increased by 1 each time the term is seen again. As long as the term does not appear in the incoming documents, we reduce its weight based on the decay function f(t). Thus, a high value of $tDF(p,t)$ corresponds to frequent terms in the recent past but not necessarily in the whole collection.

Different decay functions will handle the distribution of a term in the corpus differently. For example, a \textit{concave} decay function implies that it is more important to keep track of the number of occurrences of a term and penalize $tDF$ only when the time between two occurrences is very high. On the other hand, choosing a \textit{convex} decay function will reduce the value of $tDF$ at a higher scale when the term is absent from the near past of a news stream but then will slow down the decay until the term is absent from the sliding window. Other decay function may also make sense such as a sigmoid. We study the effect of different decay functions and discuss the results later in this paper. 

We define the extended Novelty Scoring function based on equation \ref{eq:novelty_score} as $NS^t(d,N)$:
\begin{equation}
	NS^t(d, N) = \frac{1}{norm(d)}\sum_{q \in d}{tf(q, d) \times itdf(q, N)} \label{eq:novelty_score_t}
\end{equation}
where we replace the common \textit{idf} formula with the one using the temporal document frequency ($tDF$) and that we denote by $itdf$. Note that $NS^t$ is computed based on a new document $d$ and a window size $N$ instead of a given corpus $C$ like in equation \ref{eq:novelty_score}. This is because the $tDF$ is not computed based on a corpus but it represents the \textit{recent vocabulary} of a stream.

\vspace{0.3in}

Using $tDF$ instead of $DF$ in the $IDF$ formula of equation \ref{eq:novelty_score} removes the need of keeping all documents within the sliding window. In addition, replacing the baseline document-to-document similarity measure with a term weight aggregator removes the dependence of the system complexity on the window size, making it much more computationally efficient.

%% file: 4_dataset.tex
\section{Google News Dataset}
\label{sec:GND}

As mentioned in subsection \ref{sec:relatedData}, most of the benchmark datasets used in relevant literature suffer from sparse annotations, vague assumptions and subjective judgments.
Thus, we seeked for a dataset with ground truth with regard to the first story in a news topic. Towards this direction, 
we worked for the construction of an annotated dataset from a real-world news stream that alleviates the aforementioned problems of the existing ones. 

We used the RSS feeds provided by the Google News\footnote{http://www.google.com/reader/view/} aggregator.  
The method for creating the \textit{Google news data set} was the following: we periodically collected all articles from the RSS, offered by Google News, for nine available categories. All articles are from the English news stream. Each news unit consists of the article title, a small description (snippet), the URL for the article, the publication date and a cluster id, assigned by the aggregator that cluster the news into topics. We consider an article as novel if it is the first of its cluster. In addition, we use an open source script for main content extraction from news websites \cite{Janis_2008_Content} with the article URLs to get the main content of the articles.

The final dataset used for evaluation contains articles from the category "Technology" published in the time period July 12 to August 12, 2012.  We applied standard preprocessing tasks on this data set: (a) stopword removal, for the snippet and content attributes using a predefined stopword list for the English language and (b) stemming, with the Porter stemmer. Then for each article we store the set of unigrams and their corresponding local frequency (\textit{TF}) for the article snippet and content separately.

\subsection{Annotation Process }

We take advantage of the cluster information provided by the Google News to create the ground truth dataset for our experiments. Thus, the goal set is to identify as novel, the \textit{first article} in each news cluster. Unfortunately, as the clustering in Google News is carried out via an automated mechanism, there is no guarantee that the articles in a single cluster refer to the same real world event. After analyzing the data, we concluded that the clusters correspond to topics and group together articles referring to the same theme while being close in time.

To have a reliable ground truth dataset, we assign to human annotators the task of \textit{correcting} the clusters retrieved from Google News RSS. The annotators have to assign one of the following labels to the cluster: \textit{clean}, \textit{separable}, \textit{part of an existing one} or \textit{mixed}. A clean cluster contains articles that refer to the same event (e.g. \textit{Release of iPhone5} or \textit{Ford Escape Recall}). A \textit{separable} cluster contains articles from more than one event that can easily be detected and annotated. An example of such cluster contained 22 articles for the \textit{Antitrust investigation of Microsoft by EU} and 11 articles for \textit{Windows 8 release on October 26}, two groups of articles of clearly different topics. For each \textit{separable}, the corresponding number of new clean clusters was created. When a cluster is declared as \textit{part of an existing one}, the two clusters are merged and we consider as first story the one published earlier from the union of the articles from the merged clusters. If the cluster mixes too many events that could not be easily distinguished by the annotator, the cluster is marked as \textit{mixed} and it is not considered for evaluation. We do not consider \textit{mixed} clusters for evaluation because such clusters contain more than one article that should be considered as novel. This would prompt high false alarm rates. 

The dataset we produced has some advantages over the other benchmark datasets such as TDT5.
In those datasets (some in the scale of 100,000 articles), it is the human annotators  that decide the similarity among news articles and therefore clustering before they identify the first occurrence of the cluster.
This process implies pairwise similarity evaluation by humans for the number of articles squared in the worst case!
Apparently, this process causes the introduction of noise and errors with very high probability due to the diverse background of the annotators and the chance that some articles - due to human error or negligence - are left out of the thematic news clusters. In our case the data set contains already ground truth in grouping the articles into clusters and the annotators only improve the few (compared to the documents) clusters by breaking mixed clusters into single theme ones. In any case there is no doubt on the first article per cluster as it is the temporally first in the clusters. Thus the probability for errors and more importantly missing the first article on a cluster is much smaller.

The annotation process reduced the initial data set of 3,300 articles/673 clusters to 2,006 articles/555 clusters as shown in Table \ref{tab:sizes}.

\begin{table}[ht]
\centering
\begin{tabular}{|l|c|c|} \cline{2-3}
\multicolumn{1}{c|}{}		& \textbf{\# of articles}	& \textbf{\# of clusters}\\ \hline
\textbf{Initial dataset}		& 3,300				& 673\\ \hline
\textbf{Clean dataset} 	& 2,066				 &555\\ \hline
\end{tabular}
\caption{Size of datasets used for evaluation.}
\label{tab:sizes}
\end{table}

\subsection{Dataset Characteristics}

\begin{figure}[t]
	\centering
	\includegraphics[height=5cm]{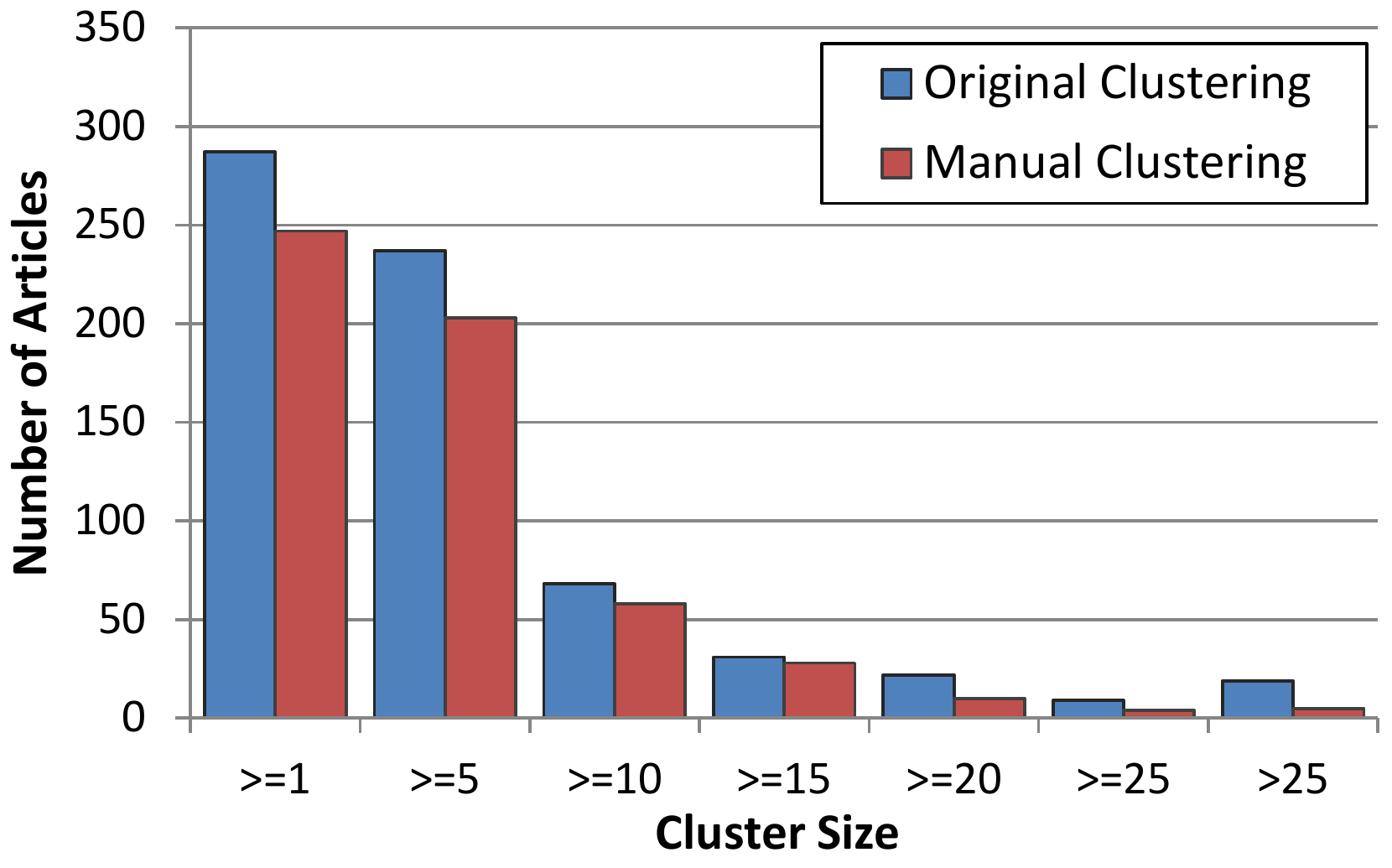}
	\caption{Distribution of cluster size in the original and final datasets.}
	\label{fig:comp_size}
\end{figure}

\begin{figure}[t]
	\centering
	\includegraphics[height=5cm]{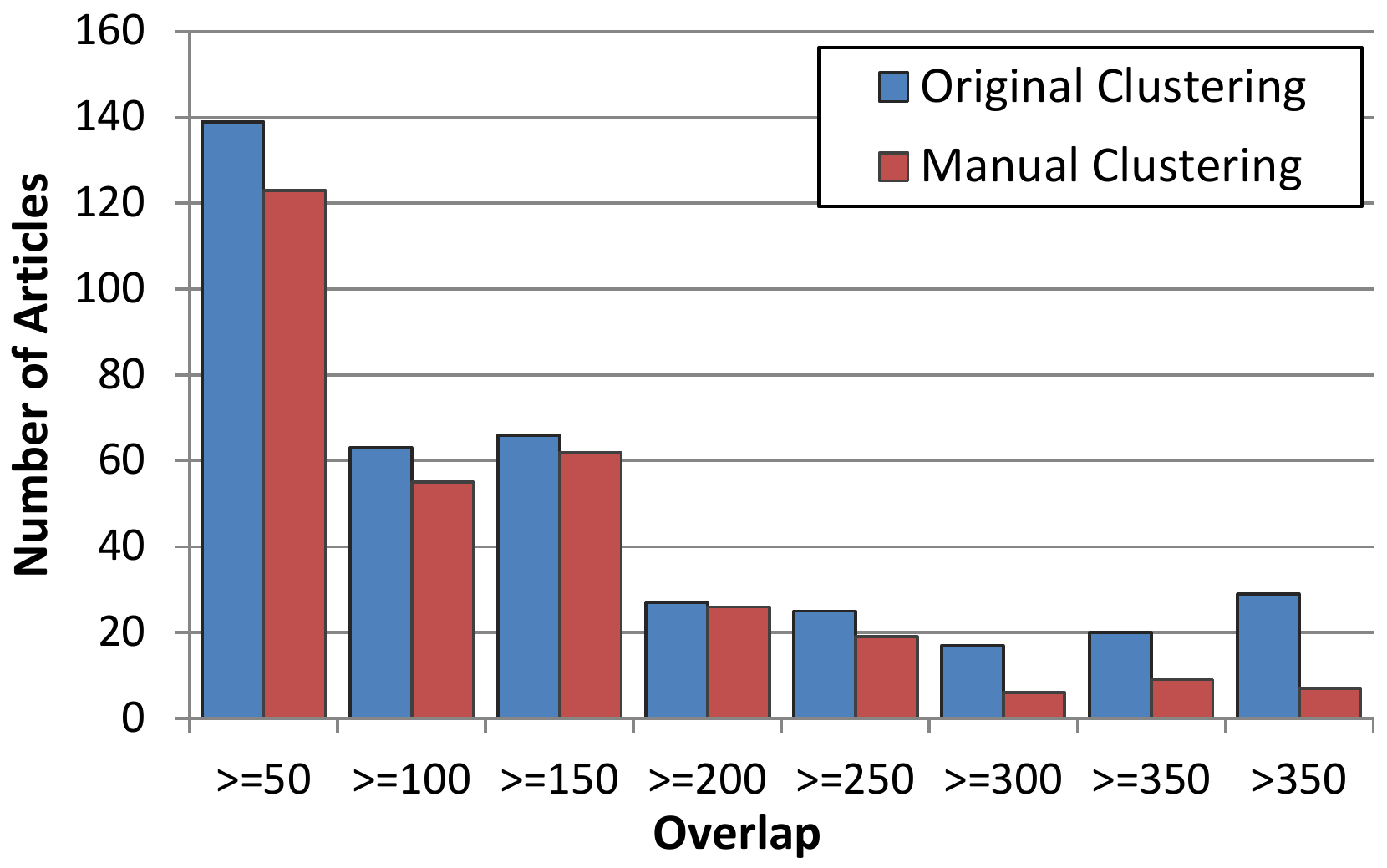}
	\caption{Distribution of overlap in the original and final datasets (number of articles from other clusters between first \& last article of a cluster.}
	\label{fig:comp_overlap}
\end{figure}

The dataset we present has two main characteristics that highlights its potential as a reliable benchmark dataset: it is an actual stream of news articles provide by a widely used news aggregator (Google News) and the annotated part corresponds to 60\% of the data. It is important to note that there is no bias in the annotated part compared to the full stream, neither in terms of cluster size nor of cluster overlap. The former is apparent in Figure \ref{fig:comp_size} where one can see the distribution of cluster size in full stream and the manually annotated part of it. The latter is shown in Figure \ref{fig:comp_overlap}. We present clusters overlap measuring the number of articles of other clusters between the first and the last article of a certain cluster.

Based on Figure \ref{fig:comp_size}, we should discuss an evident property of this stream. A large proportion of the clusters consists of only one article. The extend to this phenomenon is expected to effect the novelty detection task, positively or negatively.  Most of the articles in these clusters are guides, reviews and opinions that cannot be grouped with other articles or events of limited interest published by specialized news sites or blogs (e.g. a blog about android games reports the release of a new game application in Google Play). Clearly there should be a discussion on whether such articles should be included in a news stream intended for novelty detection. We consider the identification of different types of articles as a different problem that lays outside the field of this article. Nevertheless, we will examine the effect of the presence of such articles comparing the performance of novelty detection methods also on a subset of the dataset that exclude small clusters from evaluation (see section \ref{sec:results}). The topics of lengthy news clusters are quite characteristic -- see Table \ref{tab:cluster_sizes} for a list of the topics and sizes of the larger clusters.


\begin{table}[ht]

\centering
\begin{tabular}{|l|c|} \hline
\textbf{Cluster Topic}								&\textbf{Size}\\\hline
Skype elaborates on instant message bug				&18\\ \hline
Skype denies police surveillance policy change			&18\\ \hline
Comcast buys Microsoft stake in MSNBC.com			&19\\ \hline
Virus in Mideast spy on finance transactions			&19\\ \hline
AT\&T Unveils Shared Wireless Data Plans			&20\\ \hline
Apple Considered Investing in Twitter				&20\\ \hline
Google Nexus 7 tablet goes on sale in US				&21\\ \hline
VMware buys Nicira for \$1.05 billion					&21\\ \hline
Google unveils price for  gigabit  Internet service		&21\\ \hline
Digg acquired by Betaworks						&23\\ \hline
Microsoft Reboots Hotmail As Outlook				&27\\ \hline
FTC Fines Google for Safari Privacy Violations			&27\\ \hline
Nokia cuts Lumia 900 price in half to \$50				&30\\ \hline
Apple Brings Products Back Into EPEAT Circle			&31\\ \hline
Yahoo confirms 400k account hacks					&45\\ \hline
\end{tabular}
\caption{Sample of Topics and Cluster sizes.}
\label{tab:cluster_sizes}

\end{table} 

\begin{figure*}[t]
\begin{center}$
\begin{array}{cc}
\includegraphics[width=2.2in]{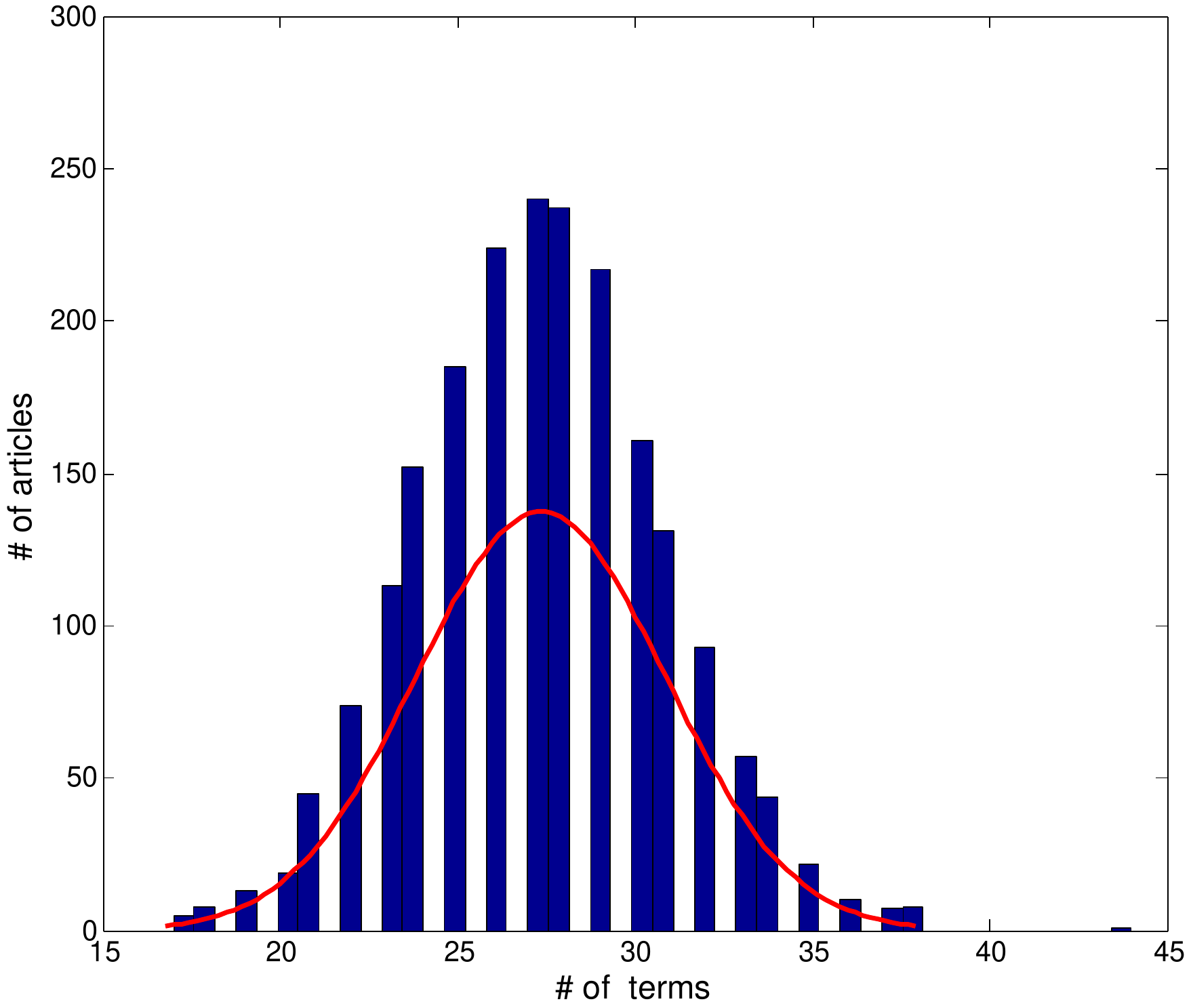} &
\includegraphics[width=2.2in]{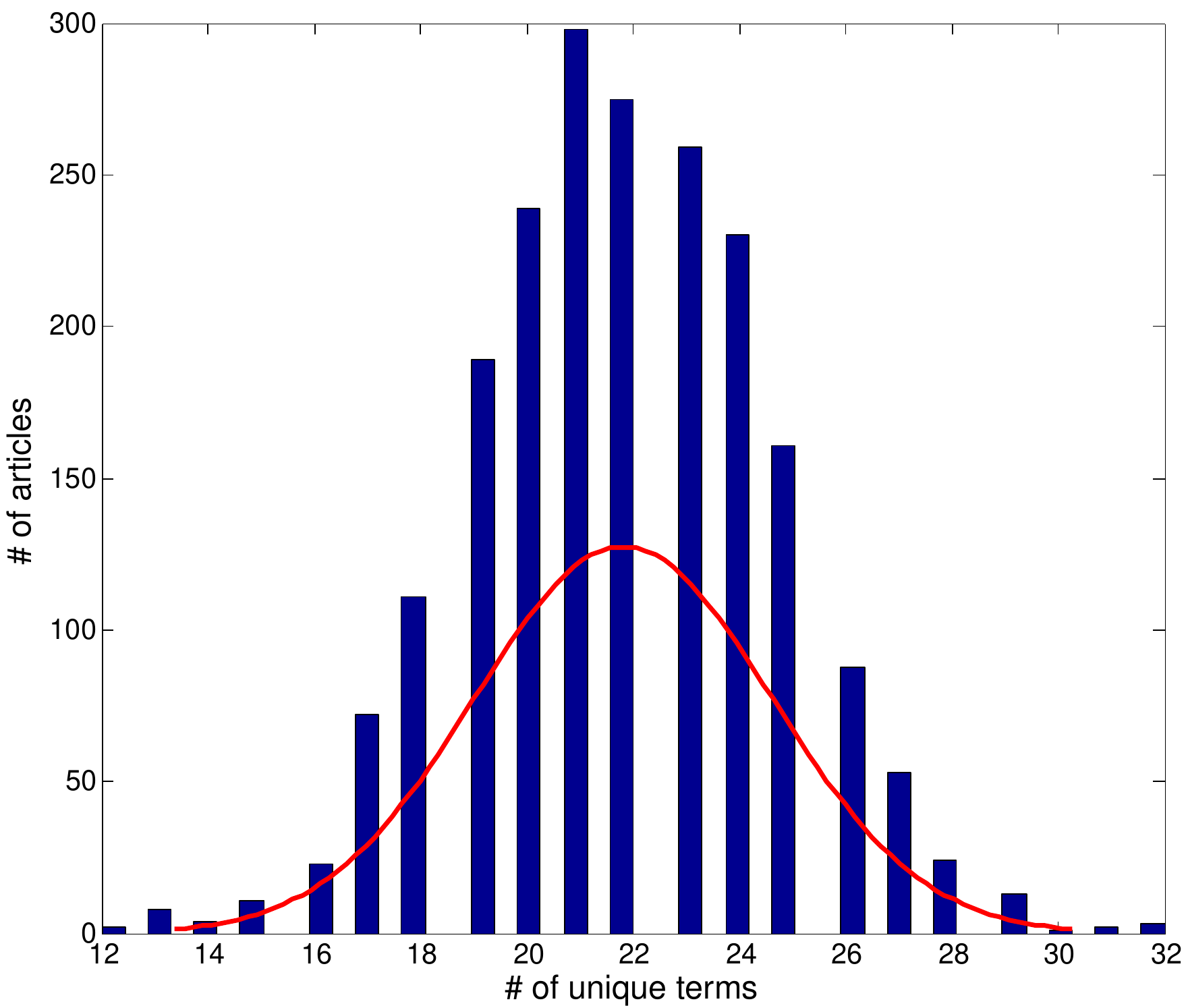} \\
\includegraphics[width=2.2in]{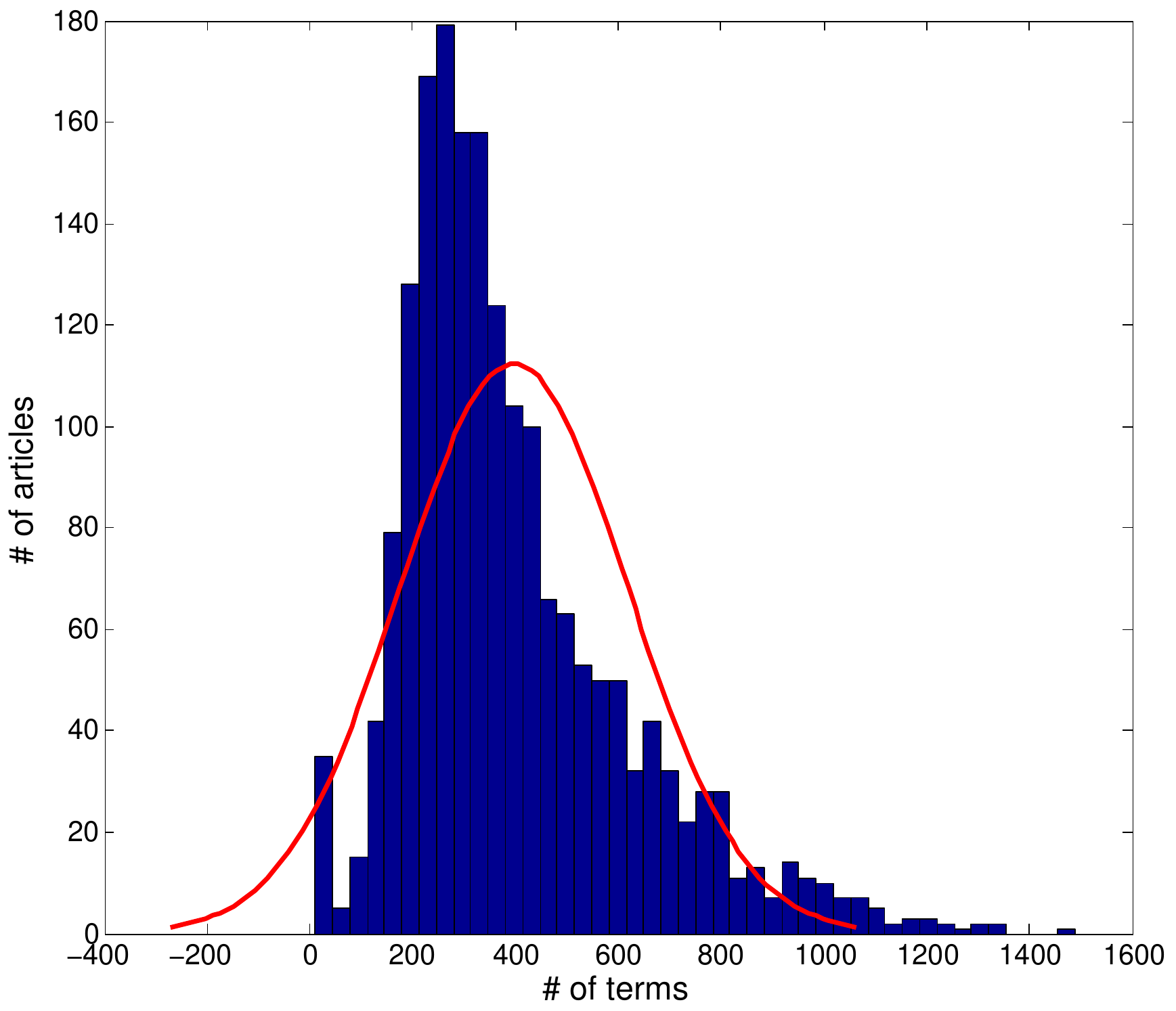} &
\includegraphics[width=2.2in]{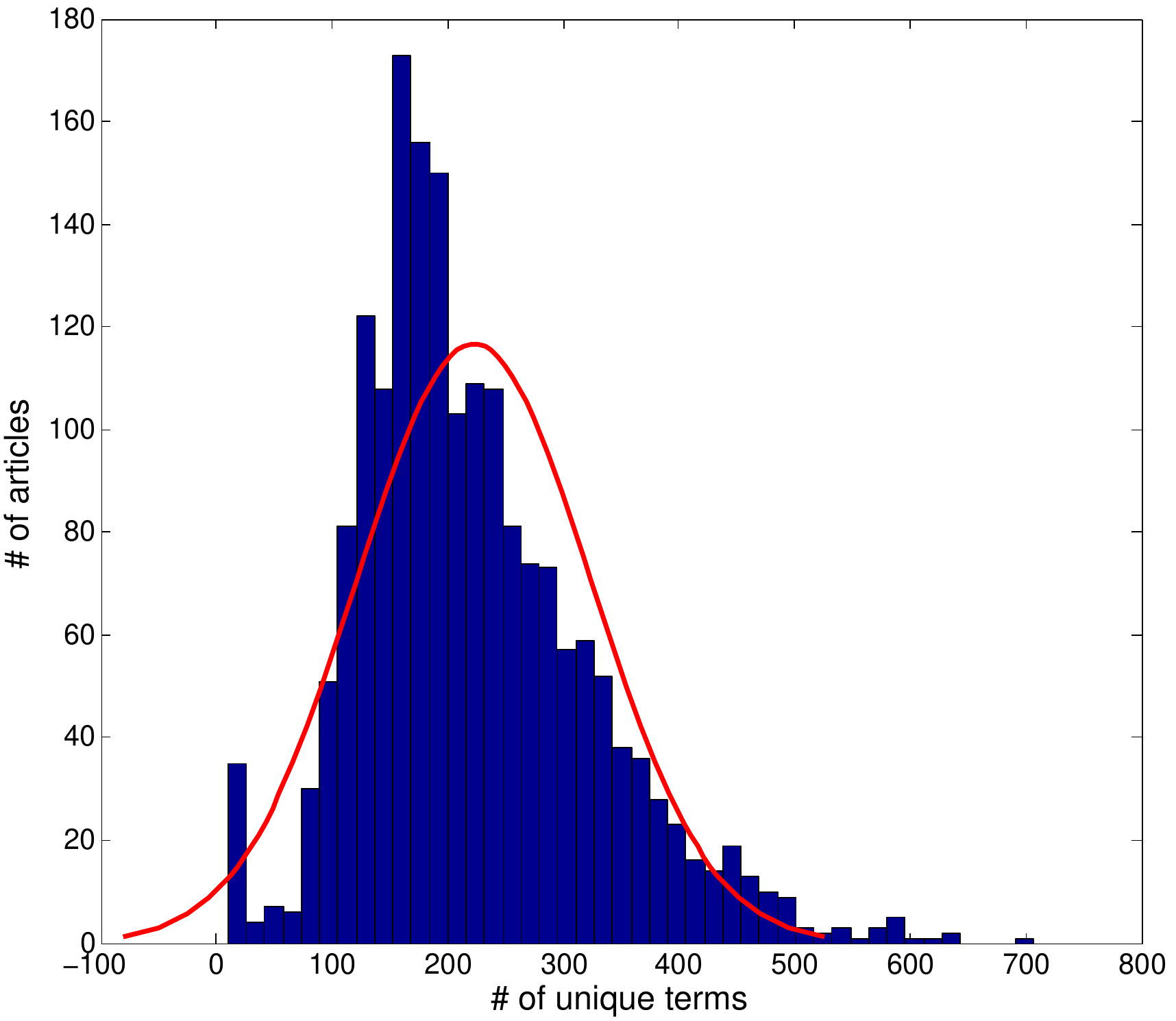}

\end{array}$
\end{center}

\caption{Distributions for number of terms in snippets (top-left), number of unique terms in snippets (top-right), number of terms in content (bottom-left) and number of unique terms in content (bottom-right).}

\label{fig:dets}
\end{figure*}

Finally, to have a complete overview of our dataset we present the distribution of all articles in terms of article length and number of unique words in Figure \ref{fig:dets} for snippets and content. Document length in snippets follows a close to normal distribution with a very small variance, around the \textit{mean of 27 terms} (or\textit{ 22 unique terms}). This is expected as Google News builds its snippets on the first few words of an article. On the other hand, when we consider the contents of articles document length varies, having a \textit{mean of 378 terms} (or \textit{209 unique terms}).

%% file: 5_experiments.tex
\section{Experimental Setting}
\label{sec:experiments}

In this section, we present the evaluation metrics, the datasets, the baselines and the various weighting models we used in our experiments.

\subsection{Evaluation Metrics}
The performance of a Novelty Detection algorithm is defined in terms of missed detection and false alarm error probabilities ($P_{Miss}$ and $P_{Fa}$) as defined in \cite{Fiscus_TDTev_2002}. 
A signal detection model, variation of ROC curves, is often used for evaluation: the Detection Error Trade-off (DET) curve \cite{Martin__det_1997}, which illustrates the trade-off between missed detections and false alarms. On the x-axis is the miss rate and on the y-axis is the false alarm rate. A system is considered to perform best when it has its curve towards the lower-left of the graph. The axis of the DET curve are on a Gaussian scale.

For the detection systems evaluation, these error probabilities are usually linearly combined into a single detection cost, $C_{Det}$, by assigning costs to missed detection and false alarm errors and specifying an a priori probability of a target. We adopt the detection cost function used in Topic Detection and Tracking (TDT) programs \cite{Fiscus_TDTev_2002,Manmatha_DCF_2002}. $C_{DET}$ is defined as follows:

\begin{equation}
	C_{Det} = C_{Miss}P_{Miss}P_{Target}+C_{Fa}P_{Fa}(1-P_{Target})
\end{equation}
where $P_{Miss}$ is the number of missed detections divided by the number of target articles, i.e. the first ones of each cluster, $P_{Fa}$ is the number of False Alarms divided by the number of non-targets, $C_{Miss}$ and $C_{Fa}$ are the costs of a missed detection and a false alarm respectively -- their values are pre-specified for the application, $P_{Miss}$ and $P_{Fa}$ are the probabilities of a missed detection and a false alarm respectively (which are determined by the evaluation results), and $P_{Target}$ is the a priori probability for finding a target as specified by the application. For our experiments we set the same cost for missed detections and false alarms ($C_{Miss}=C_{Fa}=1$) and the same probability for finding a target and a non-target ($P_{Target}=0.5$), thus there is no need for using the normalized $C_{DET}$ as described by Manmatha \textit{et al.} \citeyear{Manmatha_DCF_2002}. These are the default values for $C_{DET}$ when we assume no prior knowledge for the probability of targets.

As the goal of a detection task is to minimize both missed detections and false alarms, a detection system should minimize the detection cost $C_{DET}$. $minC_{DET}$ is used to define the optimum threshold, i.e. the threshold that gets the lowest $C_{DET}$ value and the best to use for this detection model. The  $minC_{DET}$ also corresponds to a certain point on the DET Curve, as the DET curve illustrates the different operating points of a detection system (i.e. the detection errors for different thresholds).
In order to avoid an overfitting effect over our datasets we use 5-fold cross validation in our experiments. We compute the $minC_{DET}$ on the training part and the corresponding threshold and we compute $C_{DET}$ on the testing part. We report the average detection cost for all our experiments.  

\subsection{Datasets}\label{subsec:datasets}
We conduct our experiments on the dataset described in section \ref{sec:GND}. Note that we use the actual stream including all articles published during the predefined one month period. We only exclude \textit{mixed} clusters from the final evaluation of the detection task.
We conducted experiments using both the snippets provided in RSS feeds and the main content we extracted. We use the actual stream to run the experiments and we evaluate on the annotated version of the dataset and a subset with only the clusters sized 10 articles and above.  We report the average detection cost $avgC_{DET}$ for the exhaustive combination of the above parameters and we also plot the DET curves for selected combinations of the values of different parameters.

To examine the potential of our method on very small documents, we use a second dataset consisting of actual tweets. This synthetic dataset is constructed using the annotated proportion of the one described by Petrovi\'{c} \textit{et al.} \citeyear{twitter}. The dataset contains 27 events of various lengths, from 2 to 837 tweets. The whole dataset consists of 2,600 tweets. Examples of the events includes are ``\textit{Death of Amy Winehouse}'', ``\textit{Earthquake in Virginia}'' and ``\textit{Riots break out in Tottenham}''. The stream created uses the actual temporal order of these tweets. Most of the events are well separated from each other with eight of them having a small overlap in time. Here, again we consider as novel only the first story in time for each event. The dataset is available from the website of the CROSS project\footnote{http://demeter.inf.ed.ac.uk/cross/} in the context of which it was created.
We exhaustively examine the performance of our method for the \textit{Twitter Dataset} and we present the results of the best performing weighting scheme and N value (sliding window size).

The documents in the datasets we use for experiments -- described above -- are treated as a stream of documents that need to be tagged with a novelty score upon their arrival using our novelty detection filter. The documents are sorted by publication date and for each of them a novelty score is assigned with regard to the previous $N$ documents, representing the memory of the system. We experiment with different values of $N$ in range from 20 to 200 with step 20.

\subsection{Baselines}
As mentioned earlier, we use the ground truth information of each dataset to evaluate the performance of novelty detection for our method and in comparison against the four \textit{baseline approaches} that are universally used for this task. These methods (\textit{Max Cosine Similarity}, \textit{Mean Cosine Similarity}, \textit{Max KL Divergence} and \textit{Cosine Similarity to Summary}) take into account the similarity/divergence among the document under evaluation and the previous $N$ documents or their summary and rate it as novel based on a threshold. For the experiments the weighting model used for the baselines is BM25 ($kbn$ in SMART notation -- see Table \ref{tab:smart_notations}), which is the one used by Allan \textit{et al.} \citeyear{UMass}.

\subsection{Document Weighting Models}
As mentioned in section \ref{sec:method}, we are using a variety of TF$\times$IDF weighting models (general equation \ref{eq:novelty_score}) that we will refer to using the SMART notations appearing in table \ref{tab:smart_notations}. As for $TF$, we use the variants $b$ (boolean term representation), $n$ (plain term frequency), $l$ (logarithmic concavity) and $k$ ($BM25$ saturation). 
Regarding the $IDF$ we exploit the following variants: $s$ (the plain $IDF$ value, smoothed) and $b$ (the form used in $BM25$).
Finally we consider three different options for length normalization: $n$ (none),  $d$ (the document length) and $u$ (the number of unique terms).

%% file: 6_results.tex
\section{Basic Model Performance}
\label{sec:results}

In this section we evaluate the basic novelty scoring function described in sub-section \ref{sec:NS} corresponding to equation \ref{eq:novelty_score} on both the Google News and Twitter datasets in terms of average detection cost and DET curves. We also mention the execution times for various sliding window sized compared to the baseline approaches ones.

\begin{table*}[t]
\centering
\begin{tabular}{cc}  
\includegraphics[height=11.5cm]{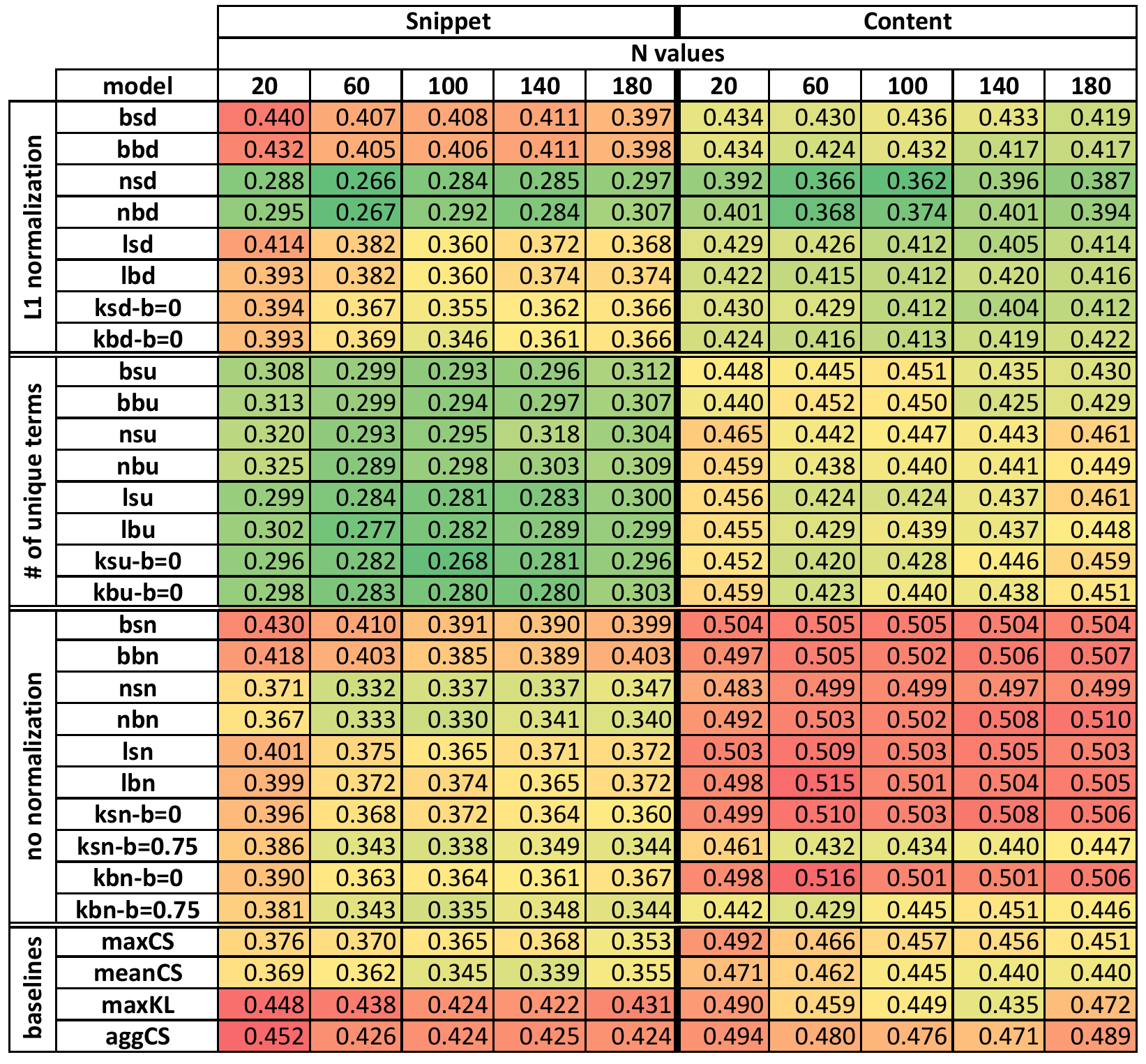} \\
\end{tabular}
\caption{Average detection cost ($avgC_{DET}$) using 5-fold cross validation on snippets and content for various TF$\times$IDF document weighting schemes compared to four baselines; the greener the better, the redder the worse.}
\label{tab:table_results}
\end{table*}

\subsection{Performance on the \textit{Google News dataset}}

\begin{figure*}[t]
\begin{center}$
\begin{array}{cc}
\includegraphics[width=2.2in]{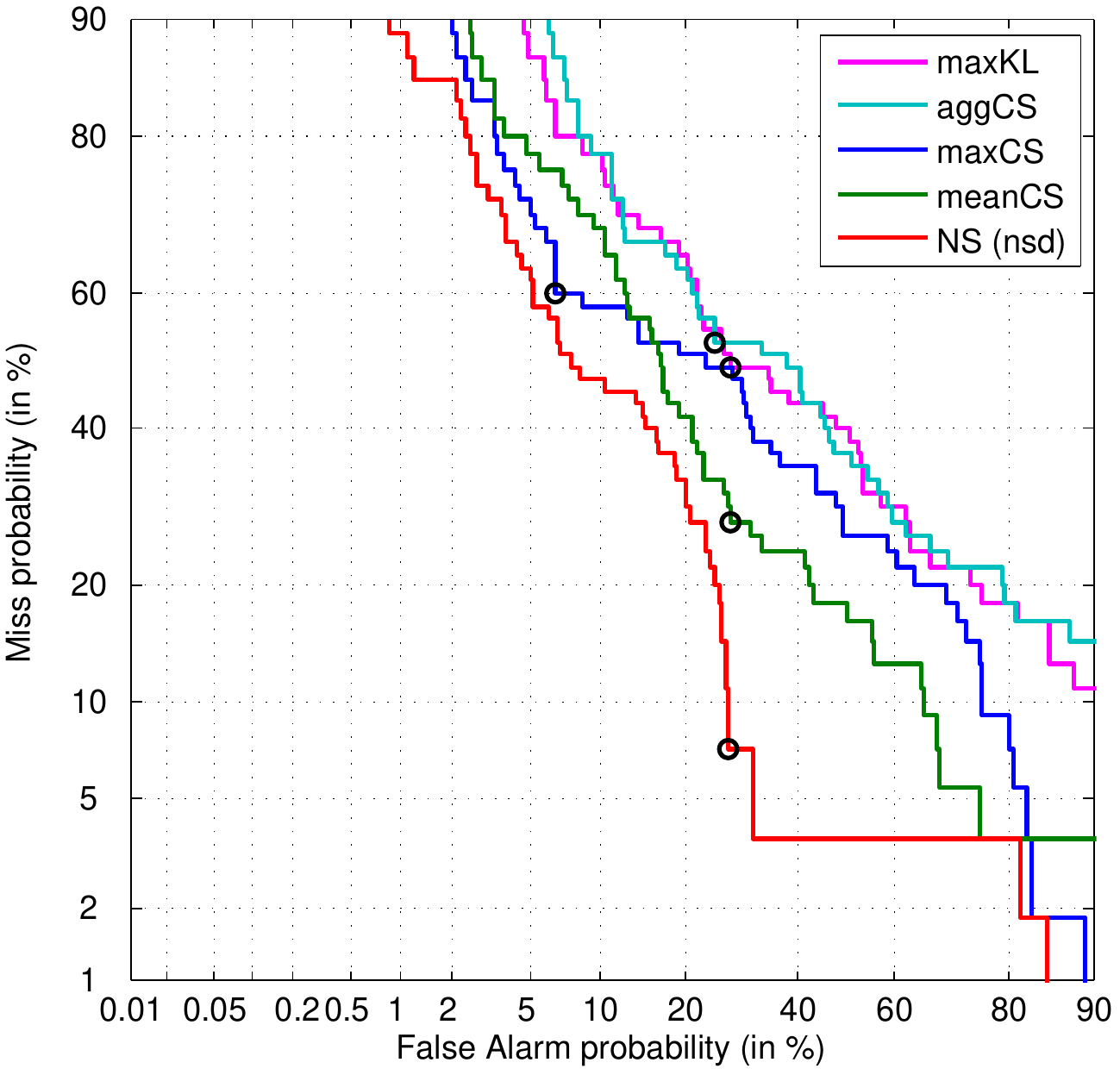} &
\includegraphics[width=2.2in]{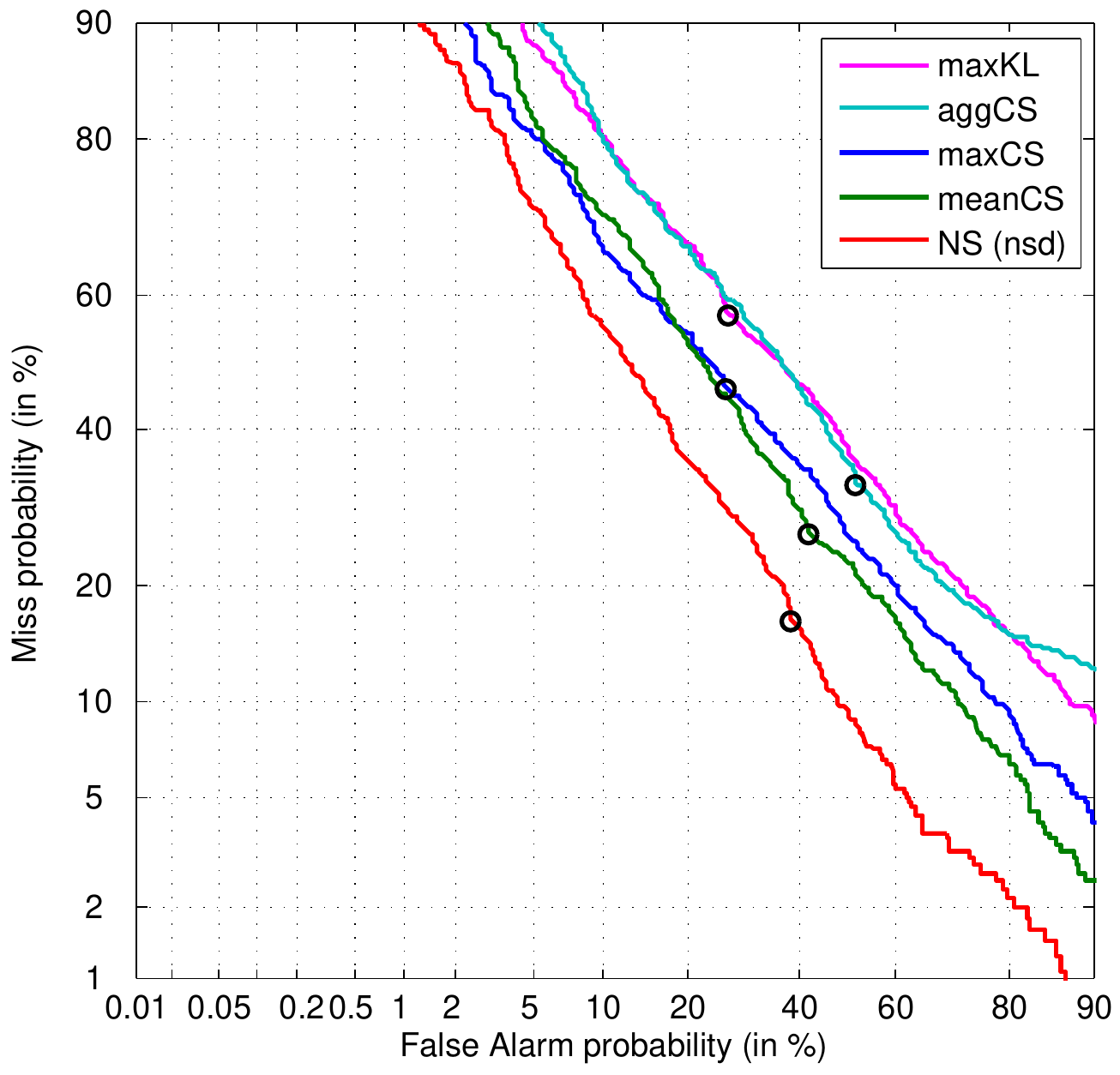} \\
\includegraphics[width=2.2in]{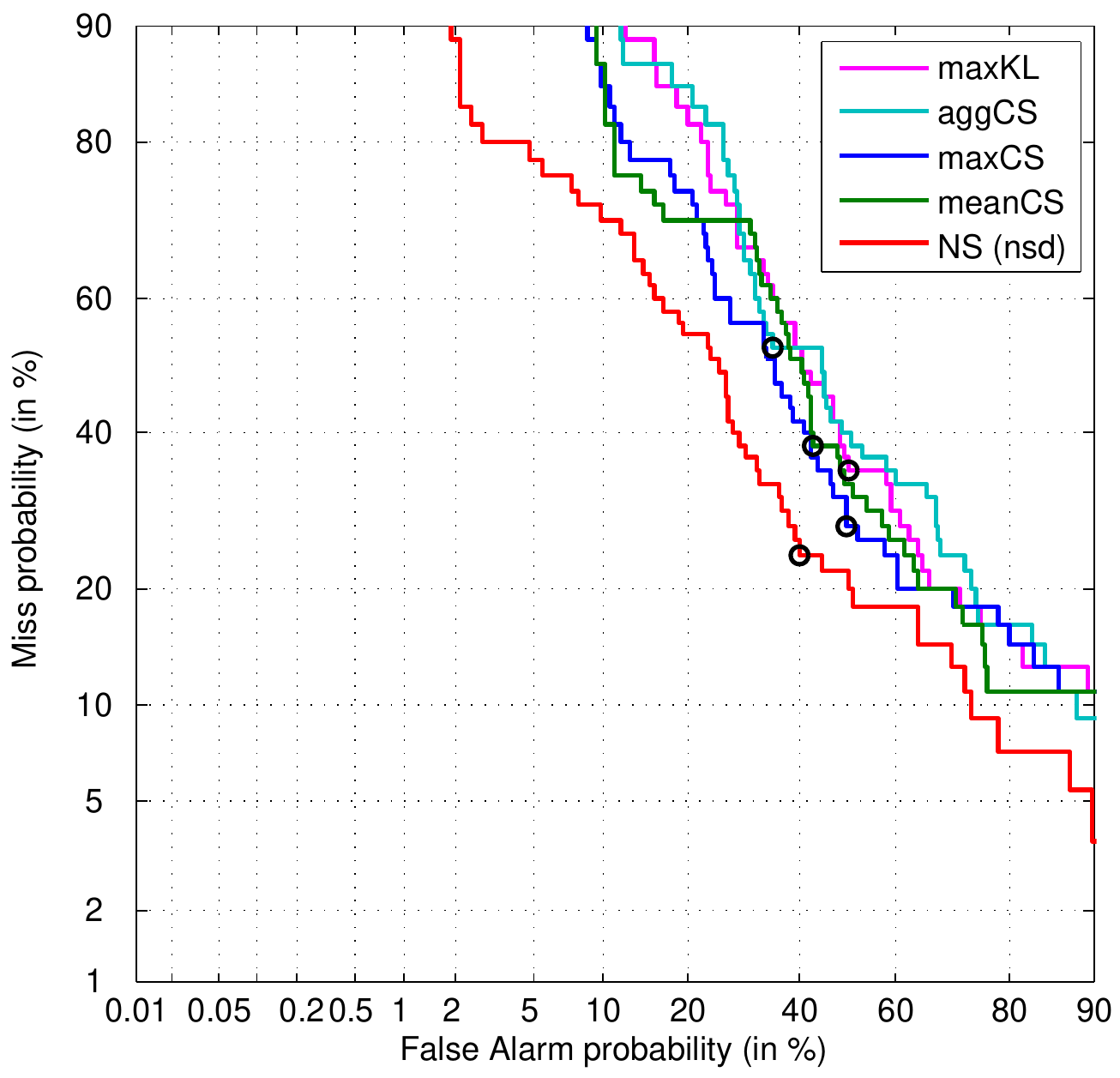} &
\includegraphics[width=2.2in]{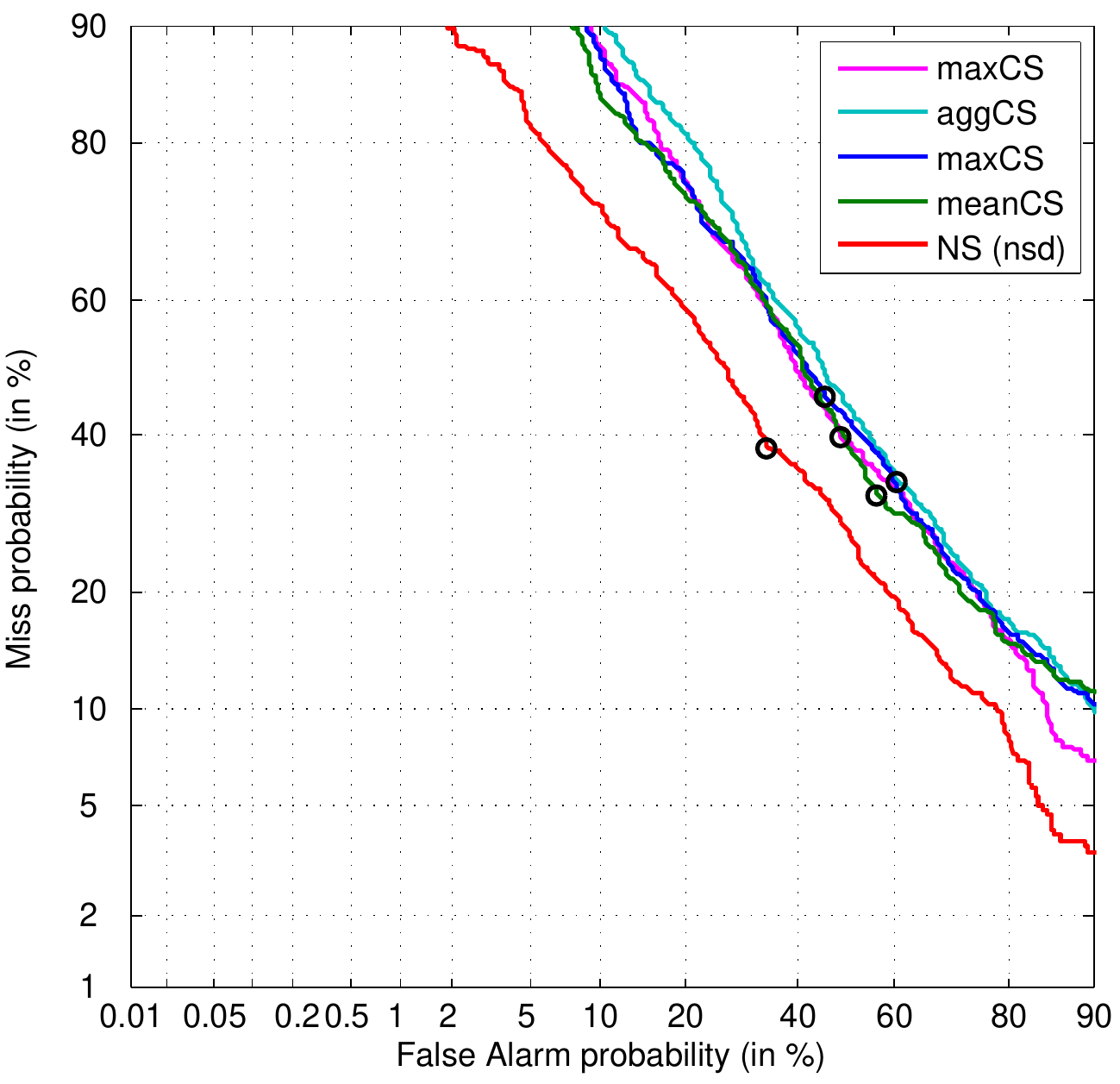}
\end{array}$
\end{center}
\caption{DET Curves for N=100 on clusters with size $\geq10$ using snippets (top-left), all clusters using snippets (top-right), clusters with size $\geq10$ using content (bottom-left) and all clusters using content (bottom-right).}
\label{fig:dets2}
\end{figure*}

In this sub-section we present and review the results of the experiments on the datasets mentioned in the previous sections and for all the combinations of measures and parameters' values mentioned for the basic model presented in subsection \ref{sec:NS}. 

\subsubsection{Average detection cost}

We present here the average detection cost ($avgC_{DET}$) for the cleaned dataset 
with memory size (i.e. length of the corpus) $N$ ranging from 20 to 200 with step 20 for a variety of meaningful combinations of the variants of term frequency, IDF and normalization. 
 
We report these results for the snippets and the full articles (Table \ref{tab:table_results}) versions of the dataset. The result table is organized in blocks of lines based on the normalization method. The top block (model SMART acronym ends with $d$) corresponds to normalization based on the document length, the mid block (SMART acronym ends with $u$) corresponds to normalization based on the number of unique terms in the document and the third one (SMART acronym ends with $n$) is for the case where no normalization takes place. The last four rows of the table represent the results of the baseline methods (\textit{MaxCS}, \textit{MeanCS}, \textit{MaxKL}, \textit{AggCS}). 

The values appearing in the cells represent the average detection cost (computed using 5-fold cross validation) for each combination of parameters. The cells' colors are based on conditional formatting where a green color indicates a low detection cost value -- the stronger the color the more favorable the result. On the contrary, the red values indicate high detection costs. The scale red to green was computed for the entire table.
We exclude some combinations of the aforementioned parameters as they are meaningless ($ksd$-$b=0.75$, $kbd$-$b=0.75$, $ksu$-$b=0.75$, $kbu$-$b=0.75$) as they introduce normalization twice. This is because TF variation $k$, used in BM25, introduces a length normalization prior to the saturation in its formula for $b>0$.

Given the above hints, we notice that almost all methods best results are obtained for memory size ($N$) either 60 or 100 thus we concentrate our further comments on the respective results columns. It is evident that the proposed Novelty Scoring measure outperforms all the baselines with the best performance achieved by a $L^1$ normalized TF-IDF (raw TF and smoothed IDF -- $nsd$) narrowly followed by the $nbd$ model (same except for the IDF component inherited from BM25). Very good performance is achieved by the u normalization (number of unique terms) especially for the $lsu$ and $kbu$ models. Absence of length normalization yields to the worst results as it can be expected with documents of varying length. Nevertheless, it still outperforms the best baseline results when we consider the snippets only (since the variation in length is limited) but performs much worse when considering the articles. The difference in performance of this group in comparison to the one on snippets originates at the greater differences in document lengths when the full article is taken into account (snippets tend to have a constant length -- around 25 terms). Note that $ksn$-$b=0.75$ and $kbn$-$b=0.75$ perform better than the rest of the block. This can be easily explained as both methods use the BM25 variant of TF which includes a pivot length normalization for parameter $b>0$. We chose to display them in that block just to be consistent in terms of SMART notations.


\subsubsection{DET Curves}

We plotted  DET curves showing the evolution of performance with regard to the Miss Detection and the False Alarm probability. These diagrams indicate the evolution of the detection cost for some of the adopted methods. They also depict the point on each curve that corresponds to the optimum threshold, having the $minC_{DET}$.  

In Figure \ref{fig:dets2}, we plot the DET Curves for memory N=100 on four versions of the Google News dataset: large clusters with size$>=10$ using snippets (top-right), size$>=10$ using content (bottom-left) and all clusters using content (bottom-right). We compare all baseline methods and our method using the best performing weighting model, $nsd$ (see Table \ref{tab:table_results}). It is clear that overall the $nsd$ method outperforms the others. The baseline based on minimum KL-divergence and document-to-summary baseline perform worst.  The same applies for the case of "all clusters" data set. 
In addition, comparing the corresponding snippet and content DET curves we confirm again our previous claims that using the full content of an article instead of a simple summary as the first few sentences of the article introduces significant noise and makes it harder to detect the first stories.

\subsection{Performance on the \textit{Twitter dataset}}

\begin{figure}[t]
\centering
\includegraphics[width=2.2in]{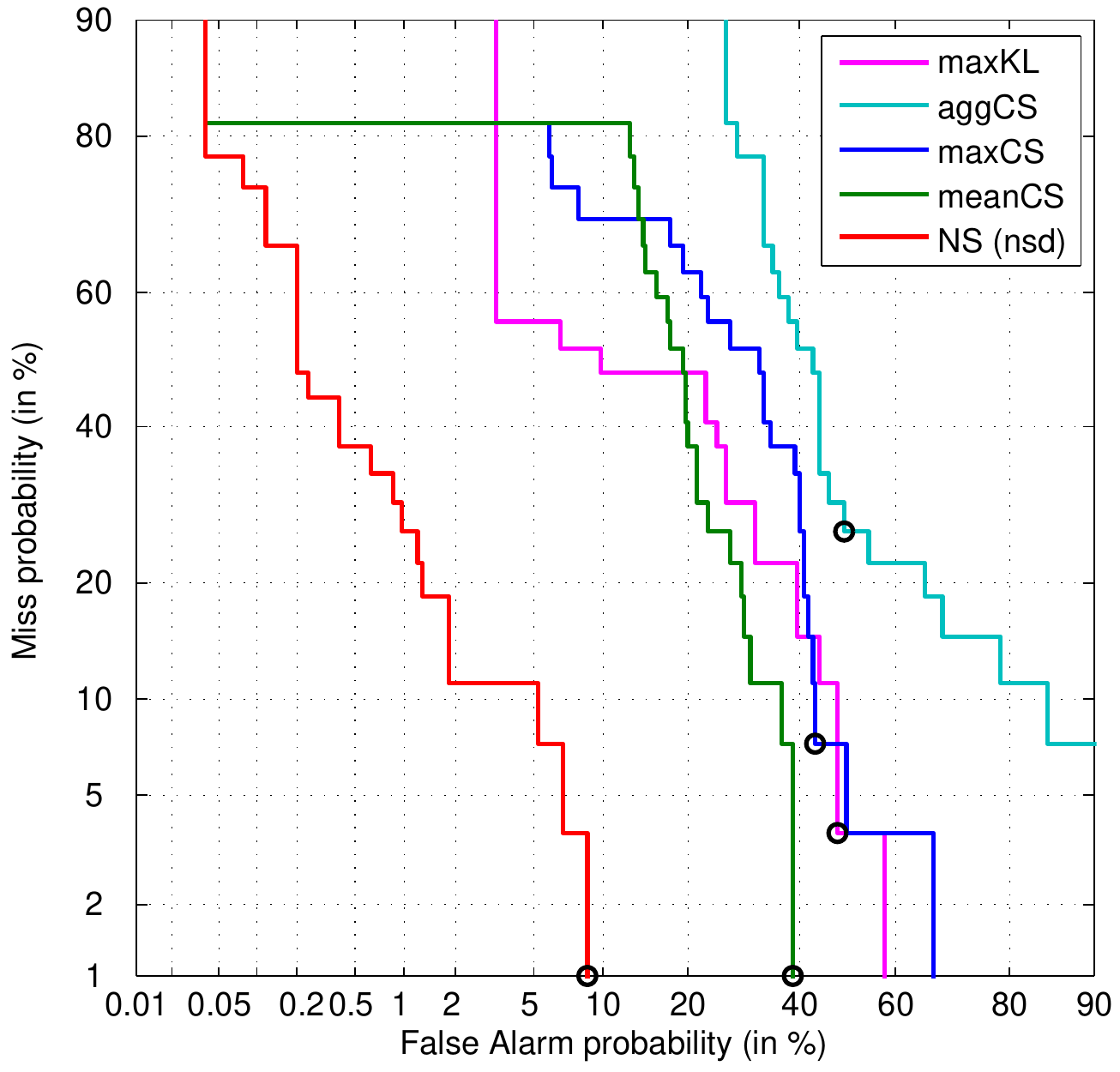}
\caption{DET Curve for N=100 on Twitter dataset.}
\label{fig:detTwitter}
\end{figure}

In Figure \ref{fig:detTwitter}, we report the results on the Twitter dataset described in section \ref{subsec:datasets}. We again compare all baseline methods and our method using the best performing weighting model, $nsd$ (see Table \ref{tab:table_results}) for $N=100$. For space constraints, we only present the results of the best performing weighting model for a given N value using the DET curves as a more concise way of presenting results. The results are very encouraging: our method outperforms by far all the baselines and manages to have a zero miss probability while maintaining false alarm probability below 10\%.

\subsection{Execution Time}

\begin{figure}[t]
\centering
\includegraphics[width=3in]{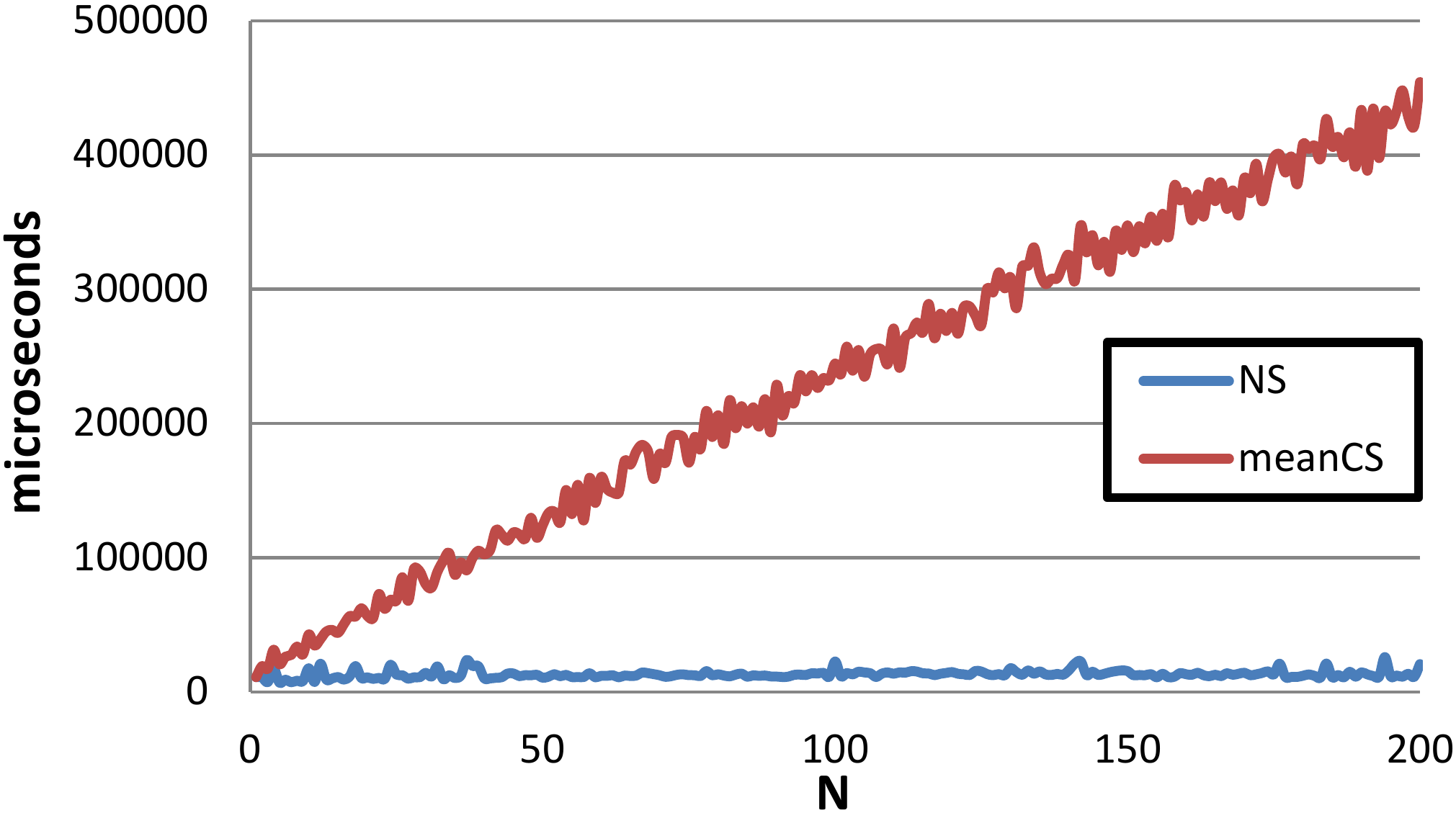}
\caption{Average execution time per document, for different values of N, for NS and meanCS.}
\label{fig:times}
\end{figure}

\begin{table*}[t]
\centering
\begin{tabular}{|r|p{40pt}|p{40pt}|p{40pt}|p{40pt}|p{40pt}|p{40pt}|p{40pt}|p{40pt}|p{40pt}|p{40pt}|}  \cline{2-6}
\multicolumn{1}{c|}{}		& \textbf{N=20}		& \textbf{N=60}		& \textbf{N=100}	& \textbf{N=140}	& \textbf{N=180}\\ \hline
\textbf{NS} 			& \textbf{124.44}	& \textbf{154.46}	& \textbf{128.50}	& \textbf{200.54}	& \textbf{134.96}\\ 
\textbf{meanCS}		& 704.06 			& 1798.30			& 2372.72 		& 3156.27 		& 3923.91\\ 
\hline
\end{tabular}
\caption{Execution times per article in microseconds for different N values, using content.}\label{tab:times}
\end{table*}

We compare our method in terms of execution time with the best performing method from the baselines, \textit{MeanCS}. 
We ran experiments for different values of $N$ using the content (and not just the snippet). We used the whole stream of news. The results are shown in Table \ref{tab:times}. The values reported correspond to the average time needed to process and assign a novelty score to an article in the dataset. The time cost for database connection and communication, indexing and index updating is not considered. The values are in microseconds.  

It is clear that our method is considerably faster than the document-to-document competing ones as it is at least seven times faster than \textit{MeanCS}. The difference among the methods increases as the corpus length increases, since \textit{MeanCS}, as any document-to-document method, have to be executed on the entire corpus to compute the similarity between all documents. 


In addition, we illustrate in Figure \ref{fig:times} the execution time in microseconds for our method and the \textit{MeanCS} baseline method, for all values of N between 1 and 200. It is evident that as the length of system memory increases, the average execution time per document keeps rising for the cosine similarity based baseline method while our method keeps having a low execution time regardless of the $N$ value.

%% file: 6.1_extended_model_evaluation.tex
\section{Extended Model Evaluation}
\label{sec:ext_mod_eval}

In this section we evaluate the extended novelty scoring function described in sub-section \ref{sec:tdf}. We adopt the best performing weighting model from the previous experiments ($nsd$) and we explore the performance of our approach for different values of the SW length ($N$), the decay parameter ($\alpha$) and a number of decay functions described below.

\subsection{Decay Functions for tDF}
\label{sec:decay_functions}

In order to evaluate the effect of employing $tDF$ in our Novelty Detection model, we compare the performance of $NS$ with and without $tDF$. For the decay functions in $tDF$, we considered four generic functions: one linear (eq. \ref{eq:linear}), one concave exponential (eq. \ref{eq:exp1}), one convex exponential (eq. \ref{eq:exp2}) and one sigmoid function (eq. \ref{eq:sig}).

\begin{align}
f_{linear}(\delta) &= 1-(\delta-1)/N \label{eq:linear}\\
f_{exp1}(\delta)  &=
  \begin{cases}
   e^{-(\delta-1)/\alpha}	& \text{if } \delta < N \\
   0       				&  \text{if } \delta = N
  \end{cases}\label{eq:exp1}\\
f_{exp2}(\delta) &= 1-e^{(\delta-N)/\alpha} \label{eq:exp2}\\
f_{sig}(\delta) &=
  \begin{cases}
   \frac{1}{{1+e^{\frac{\delta-(N/2)}{\alpha}}}}	& \text{if } \delta < N \\
   0       					&  \text{if } \delta = N
  \end{cases} \label{eq:sig}\\
\nonumber
\end{align}

In our case $\delta$ represents the distance in the past of the last occurrence of a term.
The $\alpha$ parameter for all the decay functions (except linear) controls the rate of the decay, i.e. how fast we forget a term that stops appearing in the stream.

\subsection{Performance}

We first present the results comparing the basic model with the extended when the decay function used is the \textit{linear decay}. The findings give some insights on which extent the decayed model has the potential to improve the performance of the novelty detection system.
In Table \ref{tab:linear_decay}, we present the average detection cost ($avgC_{DET}$) for basic model (no decay) and the extended model with linear decay, for different values of $N$ from 20 to 200. The best performance for both model is for $N=60$ or $N=80$.

It is apparent that depending on the evaluation set the models perform differently. For the large clusters, using snippets linear decay performs almost the same as the basic model but, for all clusters using snippets, the model with the decay outperforms the basic model by a large margin. Respectively, when using content, even though the basic model outperforms the one with linear decay, the latter performs much better when evaluated on all clusters instead of just the large ones. 

These first insights give as motivation to examine the potential of the extended model in more depth. We measure the performance of three decay functions additional to the linear decay. These decay functions, described in \ref{sec:decay_functions}, are a concave and a convex exponential and a sigmoid function. We examine them for different values of $N$, as in the previous experiments, and a number of different values of the decay parameter $\alpha$, present in all the three. The results are presented in Tables \ref{tab:alphas_ge10_snippets}, \ref{tab:alphas_all_snippets}, \ref{tab:alphas_ge10_content} and \ref{tab:alphas_all_content}, one for each evaluation set. 

We discuss the results first per decay function on the effect of parameter setting, then on differences between evaluation sets and finally on the overall comparative performance between all presented models.

\begin{table*}
\centering
\begin{tabular}{cc}
\includegraphics[height=4.5cm]{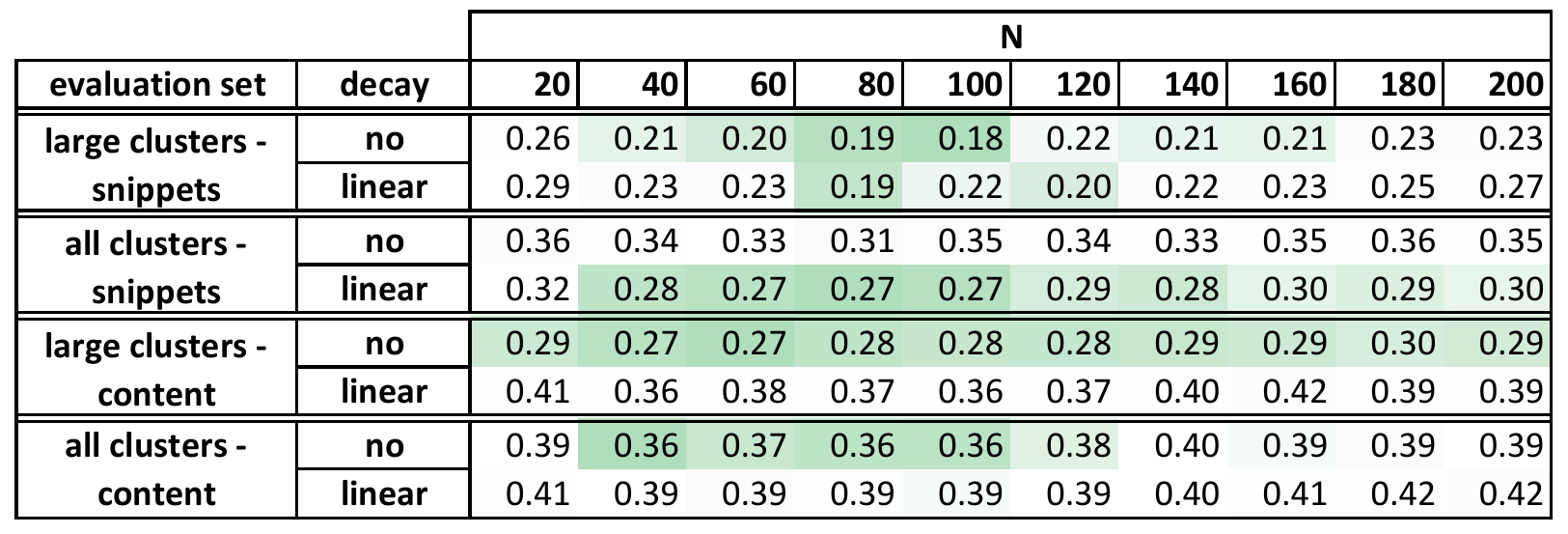} \\
\end{tabular}
\caption{Impact of the linear decay on the average detection cost; in green the best results.}
\label{tab:linear_decay}
\end{table*}

\subsubsection{Parameter Setting}

\textbf{Concave Exponential Decay (exp1)}
The first thing to observe from the four tables regarding exponential decay is the small effect of the $N$ parameter on its performance. This is expected given that the formula of concave exponential decay does not contain the parameter $N$. The effect of this parameter is to turn to zero the value of $tDF$ for the terms exiting the SW. For all evaluation sets, the optimal performance of exp1 forms a \textit{green area} around the values of $\alpha$ between 35 and 55. For smaller values of $\alpha$, i.e. when we choose to forget very fast, the performance is worse. This is because we fail to keep track of events reported by articles with a temporal distance between them. Excluding the extreme values of $N$ and $\alpha$, the performance of exponential decay is quite the same for different parameter settings, which is an important advantage of this model. Unfortunately, as we will discuss later exp1 is outperformed by all other decay functions, making it not a good choice for novelty scoring.

\textbf{Convex Exponential Decay (exp2)}
The results using exp2 stress the strong correlation between the $N$ and $\alpha$ parameters for this decay function. This correlation is translated as being the same to forget \textit{slowly} in a \textit{small} SW and to forget \textit{fast} in a \textit{large} window. Thus the \textit{red area} at the bottom left corner of the results for exp2 in all four tables corresponds in forgetting slower than needed, and a larger value of $\alpha$ should be chosen for the given value of $N$. The optimal parameter setting for this decay function is having at least $N=60$ and a large value of $\alpha=100$.

\textbf{Sigmoid Decay (sig)}
The same correlation we described for exp2 between the $N$ and $\alpha$ parameters  is apparent also for the sigmoid decay function. When the evaluation set uses snippets, the optimal results are for $N$ around 80 and $\alpha$ around 35. When the article content is used, the best results are for $N$ around 60 and $\alpha$ larger than 50.

\begin{table*}
\centering
\begin{tabular}{cc}
\includegraphics[width=9.5cm]{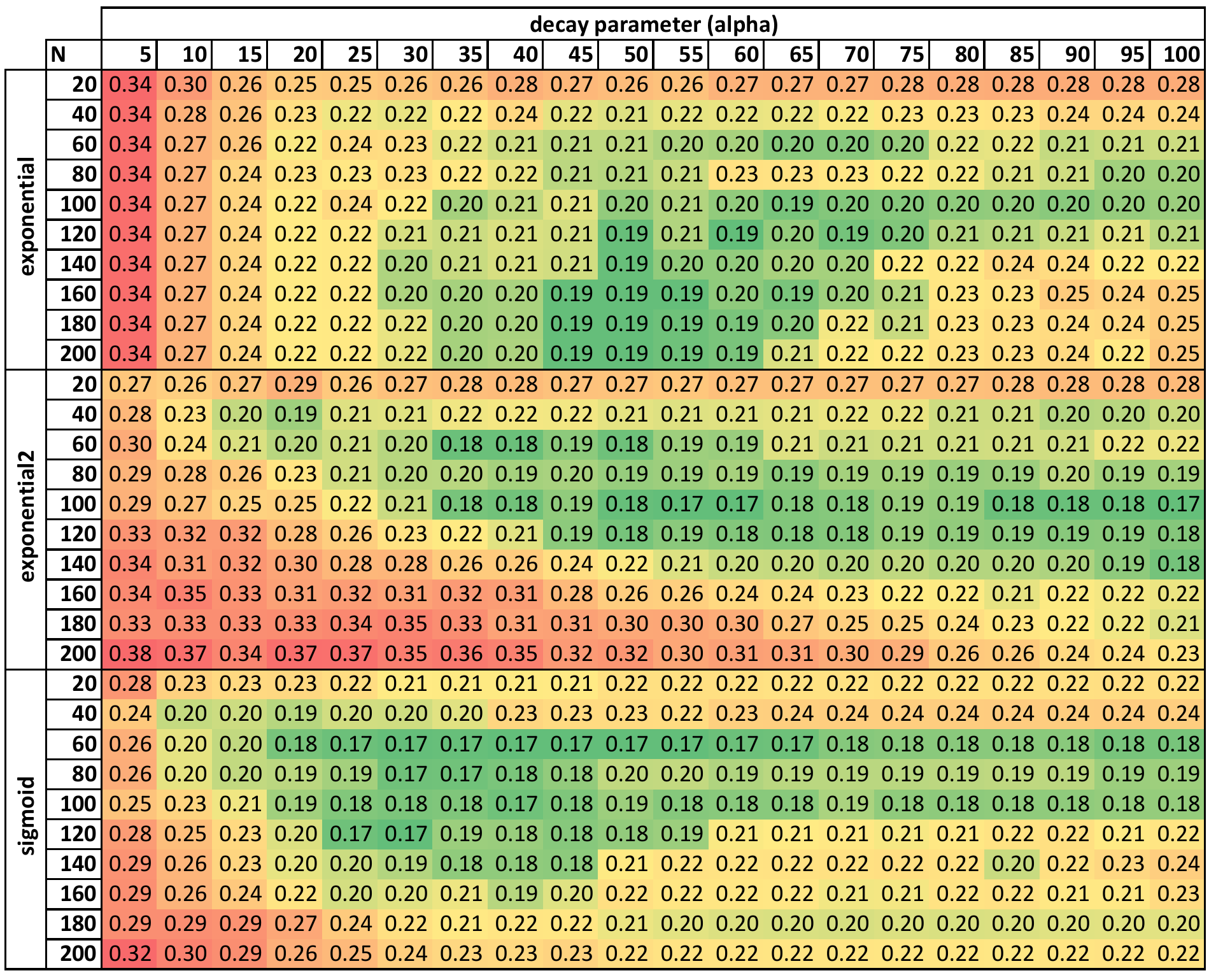} \\
\end{tabular}
\caption{Average detection cost ($avgC_{DET}$) using 5-fold cross validation on snippets of clusters of size $\geq 10$ for various decay functions and decay parameter ($\alpha$); the greener the better, the redder the worse.}
\label{tab:alphas_ge10_snippets}
\end{table*}

\begin{table*}
\centering
\begin{tabular}{cc}
\includegraphics[width=9.5cm]{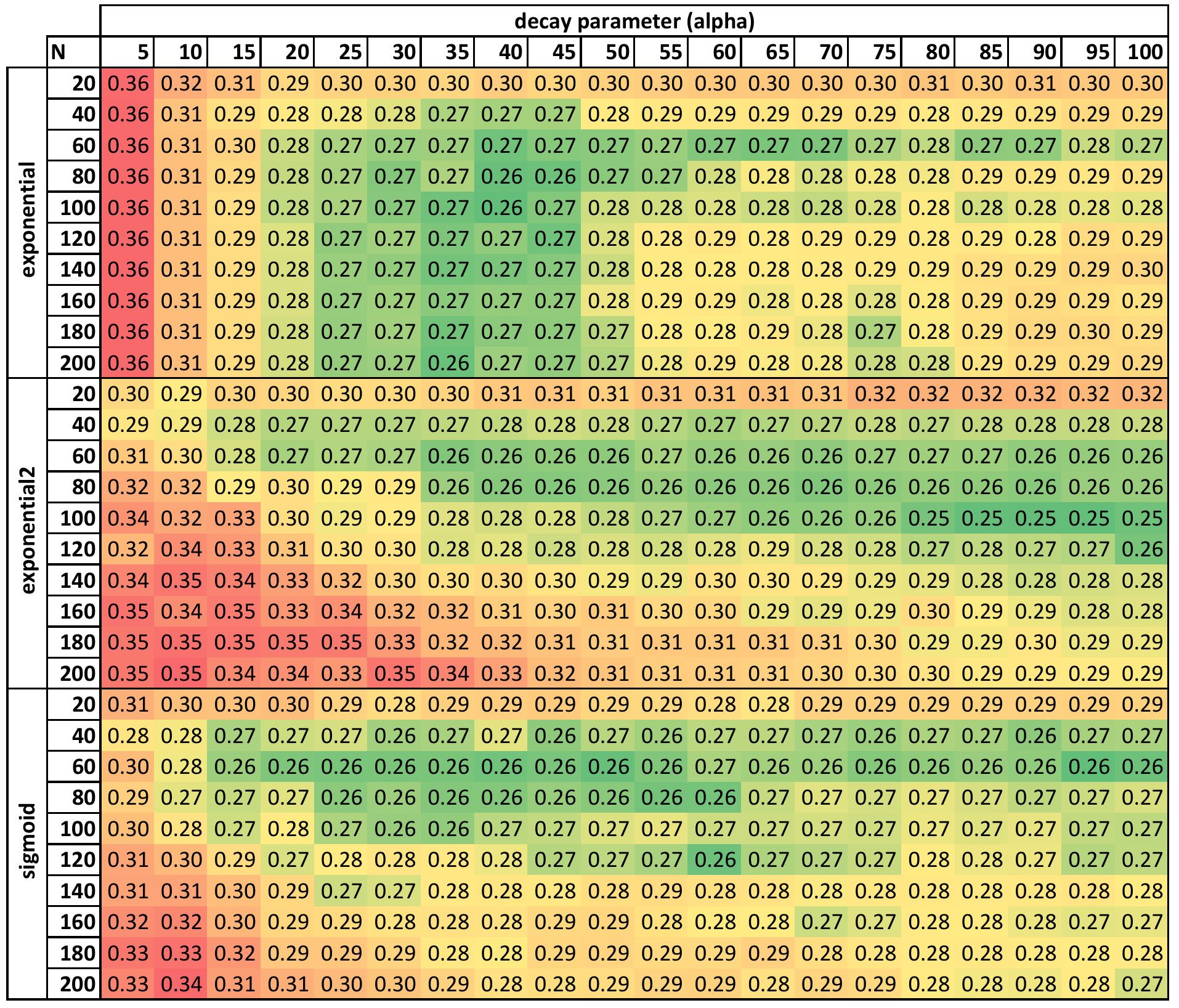} \\
\end{tabular}
\caption{Average detection cost ($avgC_{DET}$) using 5-fold cross validation on snippets of all clusters for various decay functions and decay parameter ($\alpha$); the greener the better, the redder the worse.}
\label{tab:alphas_all_snippets}
\end{table*}

\begin{table*}
\centering
\begin{tabular}{cc}
\includegraphics[width=9.5cm]{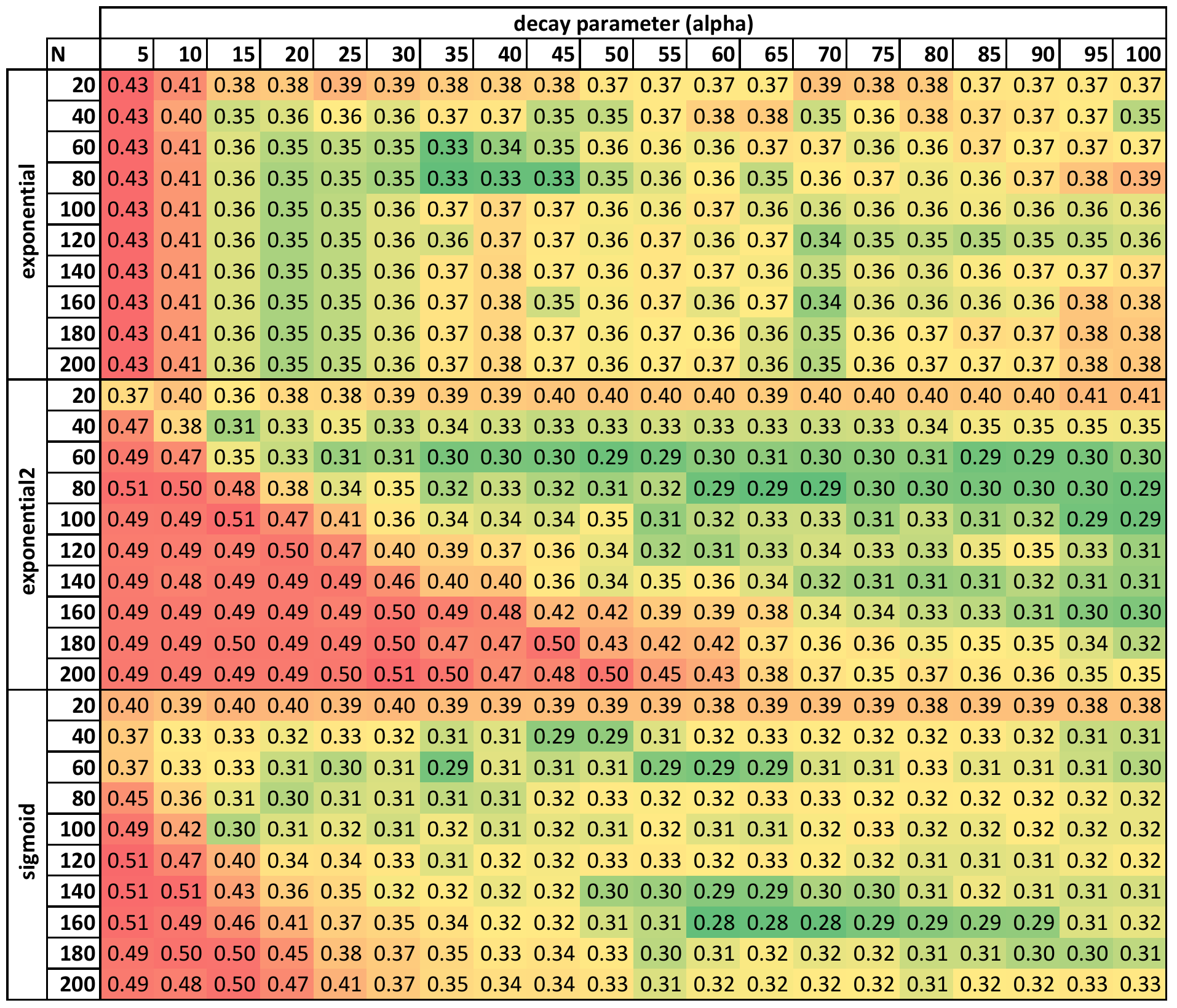} \\
\end{tabular}
\caption{Average detection cost ($avgC_{DET}$) using 5-fold cross validation on content of clusters of size $\geq 10$ for various decay functions and decay parameter ($\alpha$); the greener the better, the redder the worse.}
\label{tab:alphas_ge10_content}
\end{table*}

\begin{table*}
\centering
\begin{tabular}{cc}
\includegraphics[width=9.5cm]{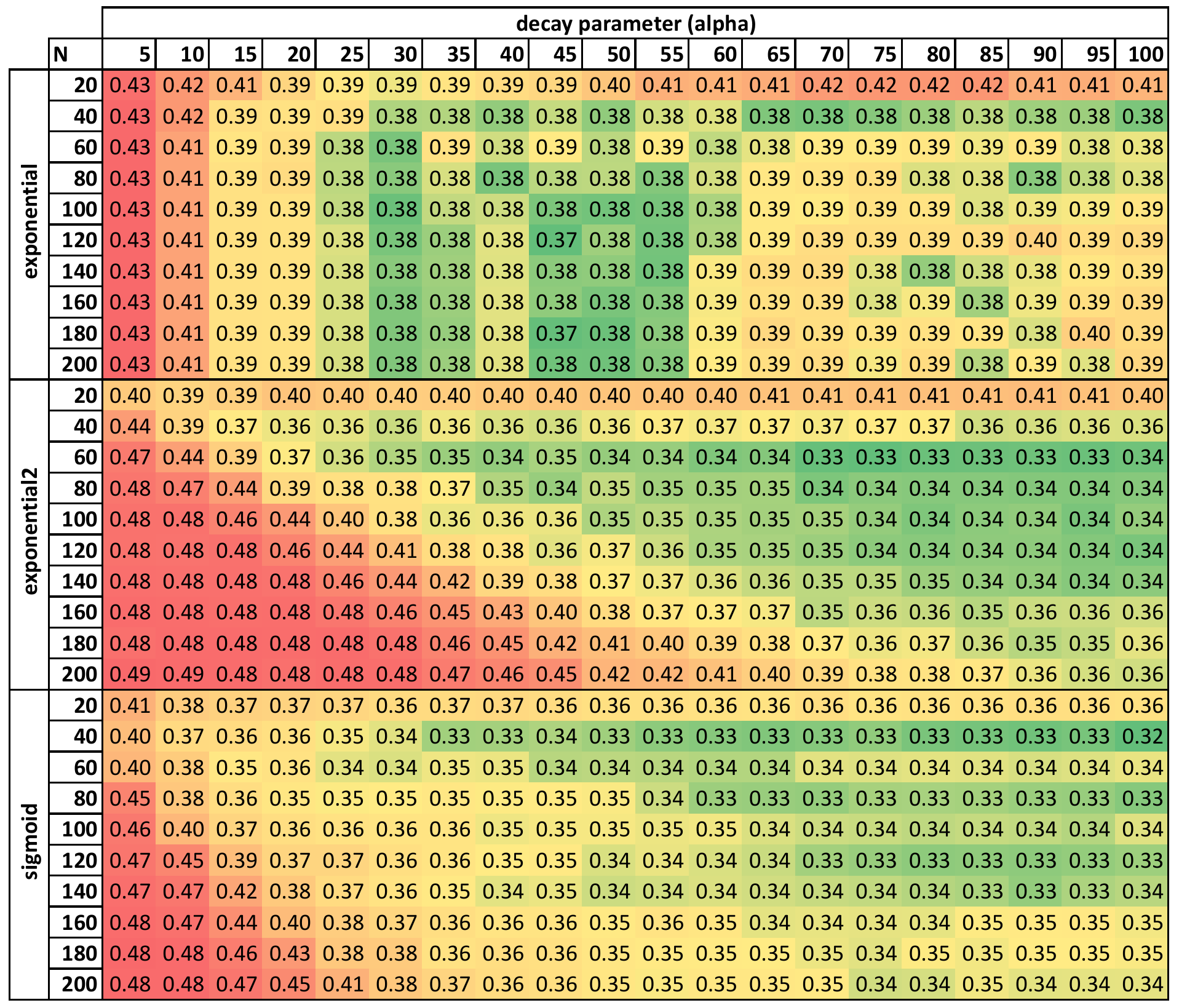} \\
\end{tabular}
\caption{Average detection cost ($avgC_{DET}$) using 5-fold cross validation on content of all clusters for various decay functions and decay parameter ($\alpha$); the greener the better, the redder the worse.}
\label{tab:alphas_all_content}
\end{table*}

\subsubsection{Evaluation Sets}

In Table \ref{tab:alphas_ge10_snippets}, we present the results for the evaluation on large clusters only using article snippet to compute its novelty score. Again, the colors are computed using conditional formatting just like in Table \ref{tab:table_results}. Taking under consideration also the results on this evaluation set from Table \ref{tab:linear_decay} for the basic model and the one using linear decay, we can see that exp2 and sigmoid decay functions outperform all the others having an $avgC_{DET}=0.17$. Despite that, note that all the models manage to have $avgC_{DET}$ lower than 0.20, showing that large clusters are easily detected using either the general or the extended model. 

When we evaluate on all clusters using the snippet (Table \ref{tab:alphas_all_snippets}) we come to an interesting finding. All decay functions have an $avgC_{DET}$ equal to 0.27 or less (including linear decay). On the contrary the basic model performs much worse, having $avgC_{DET}=0.31$. This gives a strong evidence that when the stream contains a large number of small clusters the use of the extended model boosts the performance of the novelty detection system significantly. 

When we use the article content the results change significantly. For large clusters (Table \ref{tab:alphas_ge10_content}), the best performing model is the one using the sigmoid decay ($avgC_{DET}=0.28$) with exp2 closely following ($avgC_{DET}=0.29$). Exp1 and linear decays perform worse ($avgC_{DET}=0.33$ and 0.36 respectively). In this setting, the basic model outperforms sigmoid decay model by a small margin, having a $avgC_{DET}=0.27$. Similarly, for all clusters (Table \ref{tab:alphas_all_content}), exp1 and linear decays are the worst performing ones but sigmoid outperforms both exp2 and the basic model.

\vspace{0.3in}

To conclude, we see a great potential of the extended method presented in section \ref{sec:tdf} as it outperforms the basic model in almost all evaluation settings examined in this paper. The problem pointed out when we evaluated the basic model in section \ref{sec:results} about the difficulty of detecting small clusters is alleviated through the extended model. 
The sigmoid decay function is proven to be the best performing one. This confirms a reasonable intuition that a novelty detection systems has to remember well the near past, forgetting slowly at the beginning, and forgetting faster as we get close to the end of the sliding window. This exact effect is realized using the sigmoid decay function. 

Aside from the very good performance in terms of detection accuracy, shown by the extended model, its nature allows a more efficient operation in terms of system memory required. As the index update is based on a terms last occurrence, there is no need in keeping the actual documents in our system. Thus, the extended model requires only the $tdf$ index to be stored in memory and no other information to operate.

%% file: 7_conclusions.tex
\section{Conclusion and Future Work}
\label{sec:conclusion}

Novelty detection is an important topic in modern text retrieval systems. In this paper, we proposed a new method for the novelty detection task in text streams that is more effective than several dominant baselines. We conducted extensive experiments on a real-world dataset (from a recent news stream) where our method clearly outperforms the four baseline techniques used in the relevant literature. Moreover, as our method does not use any similarity or distance measure among documents but only stream statistics kept in memory, it is much faster and scalable than the others. In addition, we examined the potential of using approximate corpus statistics for IDF alleviating the need to store the actual corpus. We incorporated the idea of temporal document frequency we introduced in \cite{tdf} for novelty scoring in order to take into account the temporal distribution of terms in the set of documents considered as corpus. The proposed extended model for novelty detection operates efficiently both in terms of execution time and system memory.

These results give strong evidence that stream statistics, such as IDF used in our method, can alone be used to detect novel documents from streams. IDF is a simple yet effective indicator of both \textit{term specificity} and \textit{document novelty}. The first property has been known since 1972 and our work just showed the second one. In large-scale streaming, such as Twitter that recently started to interest the research community, this observation may be of great importance. Alternatively, using IDF at the document level for novelty detection can be seen as a cheap estimator of the likelihood that the underlying language model for the collection of previously seen documents has generated the incoming document. Further work could try to quantify this resemblance in more depth than just comparing with the results that language models yield to as presented in this paper through one of the baseline models.

%% file: novelty_journal_paper.bbl
\begin{thebibliography}{}

\bibitem[\protect\BCAY{Allan}{Allan}{2002}]{TDT}
Allan, J. \BBOP2002\BBCP.
\newblock \BBOQ Introduction to topic detection and tracking\BBCQ\
\newblock In Allan, J.\BED, {\Bem Topic Detection and Tracking},
  \lowercase{\BVOL}~12 of {\Bem The Information Retrieval Series}, \BPGS\
  1--16. Springer US.

\bibitem[\protect\BCAY{Allan, Lavrenko,\ \BBA\ Jin}{Allan
  et~al.}{2000a}]{Allan_Hard_2000}
Allan, J., Lavrenko, V., \BBA\ Jin, H. \BBOP2000a\BBCP.
\newblock \BBOQ First story detection in tdt is hard\BBCQ\
\newblock In {\Bem 9th international Conference on Information and Knowledge
  Management}, CIKM '00, \BPGS\ 374--381. ACM.

\bibitem[\protect\BCAY{Allan, Lavrenko, Malin,\ \BBA\ Swan}{Allan
  et~al.}{2000b}]{UMass}
Allan, J., Lavrenko, V., Malin, D., \BBA\ Swan, R. \BBOP2000b\BBCP.
\newblock \BBOQ Detections, bounds, and timelines: Umass and tdt-3\BBCQ\
\newblock In {\Bem Topic Detection and Tracking Workshop (TDT-3)}.

\bibitem[\protect\BCAY{Allan, Wade,\ \BBA\ Bolivar}{Allan
  et~al.}{2003}]{sentence}
Allan, J., Wade, C., \BBA\ Bolivar, A. \BBOP2003\BBCP.
\newblock \BBOQ Retrieval and novelty detection at the sentence level\BBCQ\
\newblock In {\Bem 26th annual international ACM SIGIR conference on Research
  and development in information retrieval}, SIGIR '03, \BPGS\ 314--321. ACM.

\bibitem[\protect\BCAY{Elsts}{Elsts}{2008}]{Janis_2008_Content}
Elsts, J. \BBOP2008\BBCP.
\newblock \BBOQ Extracting the main content from a webpage\BBCQ.

\bibitem[\protect\BCAY{Fang, Tao,\ \BBA\ Zhai}{Fang
  et~al.}{2004}]{fang_formal_2004}
Fang, H., Tao, T., \BBA\ Zhai, C. \BBOP2004\BBCP.
\newblock \BBOQ A formal study of information retrieval heuristics\BBCQ\
\newblock In {\Bem 27th annual international ACM SIGIR conference on Research
  and development in information retrieval}, {SIGIR} '04, \BPGS\ 49--56. {ACM}.

\bibitem[\protect\BCAY{Fiscus\ \BBA\ Doddington}{Fiscus\ \BBA\
  Doddington}{2002}]{Fiscus_TDTev_2002}
Fiscus, J.~G.\BBACOMMA\  \BBA\ Doddington, G.~R. \BBOP2002\BBCP.
\newblock \BBOQ Topic detection and tracking\BBCQ\
\newblock In Allan, J.\BED, {\Bem Topic detection and tracking}, \BCH~1, \BPGS\
  17--31. Kluwer Academic Publishers.

\bibitem[\protect\BCAY{Harman}{Harman}{2002}]{trec2002}
Harman, D. \BBOP2002\BBCP.
\newblock \BBOQ Overview of the trec 2002 novelty track\BBCQ\
\newblock In {\Bem 11th Text REtrieval Conference (TREC 2002), NIST Special
  Publication 500-251}, \BPGS\ 46--55.

\bibitem[\protect\BCAY{Karkali, Plachouras, Stefanatos,\ \BBA\
  Vazirgiannis}{Karkali et~al.}{2012}]{tdf}
Karkali, M., Plachouras, V., Stefanatos, C., \BBA\ Vazirgiannis, M.
  \BBOP2012\BBCP.
\newblock \BBOQ Keeping keywords fresh: a bm25 variation for personalized
  keyword extraction\BBCQ\
\newblock In {\Bem 2nd Temporal Web Analytics Workshop}, TempWeb '12, \BPGS\
  17--24. ACM.

\bibitem[\protect\BCAY{Kwee, Tsai,\ \BBA\ Tang}{Kwee et~al.}{2009}]{sl2}
Kwee, A.~T., Tsai, F.~S., \BBA\ Tang, W. \BBOP2009\BBCP.
\newblock \BBOQ Sentence-level novelty detection in english and malay\BBCQ\
\newblock In {\Bem 13th Pacific-Asia Conference on Advances in Knowledge
  Discovery and Data Mining}, PAKDD '09, \BPGS\ 40--51. Springer-Verlag.

\bibitem[\protect\BCAY{Li\ \BBA\ Croft}{Li\ \BBA\ Croft}{2005}]{sl1}
Li, X.\BBACOMMA\  \BBA\ Croft, W.~B. \BBOP2005\BBCP.
\newblock \BBOQ Novelty detection based on sentence level patterns\BBCQ\
\newblock In {\Bem 14th ACM international conference on Information and
  knowledge management}, CIKM '05, \BPGS\ 744--751. ACM.

\bibitem[\protect\BCAY{Lin\ \BBA\ Brusilovsky}{Lin\ \BBA\
  Brusilovsky}{2011}]{hypermedia}
Lin, Y.-l.\BBACOMMA\  \BBA\ Brusilovsky, P. \BBOP2011\BBCP.
\newblock \BBOQ Towards open corpus adaptive hypermedia: a study of novelty
  detection approaches\BBCQ\
\newblock In {\Bem 19th international conference on User modeling, adaption,
  and personalization}, UMAP'11, \BPGS\ 353--358. Springer-Verlag.

\bibitem[\protect\BCAY{Luo, Tang,\ \BBA\ Yu}{Luo et~al.}{2007}]{Luo_TED_2007}
Luo, G., Tang, C., \BBA\ Yu, P.~S. \BBOP2007\BBCP.
\newblock \BBOQ Resource-adaptive real-time new event detection\BBCQ\
\newblock In {\Bem 2007 ACM SIGMOD international conference on Management of
  data}, SIGMOD '07, \BPGS\ 497--508. ACM.

\bibitem[\protect\BCAY{Manmatha, Feng,\ \BBA\ Allan}{Manmatha
  et~al.}{2002}]{Manmatha_DCF_2002}
Manmatha, R., Feng, A., \BBA\ Allan, J. \BBOP2002\BBCP.
\newblock \BBOQ A critical examination of tdt's cost function\BBCQ\
\newblock In {\Bem 25th annual international ACM SIGIR conference on Research
  and development in information retrieval}, SIGIR '02, \BPGS\ 403--404. ACM.

\bibitem[\protect\BCAY{Markou\ \BBA\ Singh}{Markou\ \BBA\ Singh}{2003a}]{part1}
Markou, M.\BBACOMMA\  \BBA\ Singh, S. \BBOP2003a\BBCP.
\newblock \BBOQ Novelty detection a review--part 1: statistical
  approaches\BBCQ\
\newblock {\Bem Signal Process.}, {\Bem 83\/}(12), 2481--2497.

\bibitem[\protect\BCAY{Markou\ \BBA\ Singh}{Markou\ \BBA\ Singh}{2003b}]{part2}
Markou, M.\BBACOMMA\  \BBA\ Singh, S. \BBOP2003b\BBCP.
\newblock \BBOQ Novelty detection a review-part 2: neural network based
  approaches\BBCQ\
\newblock {\Bem Signal Processing}, {\Bem 83\/}(12), 2499--2521.

\bibitem[\protect\BCAY{Martin, Doddington, Kamm, Ordowski,\ \BBA\
  Przybocki}{Martin et~al.}{1997}]{Martin__det_1997}
Martin, A., Doddington, G., Kamm, T., Ordowski, M., \BBA\ Przybocki, M.
  \BBOP1997\BBCP.
\newblock \BBOQ The det curve in assessment of detection task performance\BBCQ\
\newblock In {\Bem 5th European Conference on Speech Communication and
  Technology}, \BPGS\ 1895--1898.

\bibitem[\protect\BCAY{Ohgaya, Shimmura, Takagi,\ \BBA\ Aizawa}{Ohgaya
  et~al.}{2003}]{Meiji_2003}
Ohgaya, R., Shimmura, A., Takagi, T., \BBA\ Aizawa, A.~N. \BBOP2003\BBCP.
\newblock \BBOQ Meiji university web and novelty track experiments at trec
  2003\BBCQ\
\newblock In {\Bem The Twelth Text Retrieval Conference (TREC)}, \BPGS\
  399--407.

\bibitem[\protect\BCAY{Petrovi\'{c}, Osborne,\ \BBA\ Lavrenko}{Petrovi\'{c}
  et~al.}{2010}]{twitter}
Petrovi\'{c}, S., Osborne, M., \BBA\ Lavrenko, V. \BBOP2010\BBCP.
\newblock \BBOQ Streaming first story detection with application to
  twitter\BBCQ\
\newblock In {\Bem Human Language Technologies: The 2010 Annual Conference of
  the North American Chapter of the Association for Computational Linguistics},
  HLT '10, \BPGS\ 181--189. ACL.

\bibitem[\protect\BCAY{Robertson\ \BBA\ Walker}{Robertson\ \BBA\
  Walker}{1997}]{robertson_relevance_1997}
Robertson, S.~E.\BBACOMMA\  \BBA\ Walker, S. \BBOP1997\BBCP.
\newblock \BBOQ On relevance weights with little relevance information\BBCQ\
\newblock {\Bem {SIGIR} Forum}, {\Bem 31\/}({SI}), 16--24.

\bibitem[\protect\BCAY{Robertson, Walker, Sparck~Jones, Hancock-Beaulieu,\
  \BBA\ Gatford}{Robertson et~al.}{1994}]{robertson_okapi_1994}
Robertson, S.~E., Walker, S., Sparck~Jones, K., Hancock-Beaulieu, M., \BBA\
  Gatford, M. \BBOP1994\BBCP.
\newblock \BBOQ Okapi at {TREC-3}\BBCQ\
\newblock In {\Bem 3rd Text {REtrieval} Conference}, {TREC-3}, \BPGS\ 109--126.

\bibitem[\protect\BCAY{Singhal, Salton,\ \BBA\ Buckley}{Singhal
  et~al.}{1995}]{singhal_length_1995}
Singhal, A., Salton, G., \BBA\ Buckley, C. \BBOP1995\BBCP.
\newblock \BBOQ Length normalization in degraded text collections\BBCQ\
\newblock \BTR, Cornell University, Ithaca, {NY}, {USA}.

\bibitem[\protect\BCAY{Soboroff}{Soboroff}{2004}]{trec2004}
Soboroff, I. \BBOP2004\BBCP.
\newblock \BBOQ Overview of the trec 2004 novelty track\BBCQ\
\newblock In {\Bem 13th Text REtrieval Conference (TREC 2004), NIST Special
  Publication 500-251}.

\bibitem[\protect\BCAY{Soboroff\ \BBA\ Harman}{Soboroff\ \BBA\
  Harman}{2003}]{trec2003}
Soboroff, I.\BBACOMMA\  \BBA\ Harman, D. \BBOP2003\BBCP.
\newblock \BBOQ Overview of the trec 2003 novelty track\BBCQ\
\newblock In {\Bem 12th Text REtrieval Conference (TREC 2003), NIST Special
  Publication 500-251}.

\bibitem[\protect\BCAY{Soboroff\ \BBA\ Harman}{Soboroff\ \BBA\
  Harman}{2005}]{TREC}
Soboroff, I.\BBACOMMA\  \BBA\ Harman, D. \BBOP2005\BBCP.
\newblock \BBOQ Novelty detection: the trec experience\BBCQ\
\newblock In {\Bem Conference on Human Language Technology and Empirical
  Methods in Natural Language Processing}, HLT '05, \BPGS\ 105--112. ACL.

\bibitem[\protect\BCAY{Sparck~Jones}{Sparck~Jones}{1972}]{sparck_jones_statistical_1972}
Sparck~Jones, K. \BBOP1972\BBCP.
\newblock \BBOQ A statistical interpretation of term specificity and its
  application in retrieval\BBCQ\
\newblock {\Bem Journal of Documentation}, {\Bem 28\/}(1), 11--20.

\bibitem[\protect\BCAY{Tsai}{Tsai}{2010}]{Tsai_Review_2010}
Tsai, F.~S. \BBOP2010\BBCP.
\newblock \BBOQ Review of techniques for intelligent novelty mining\BBCQ\
\newblock {\Bem Information Technology Journal}, {\Bem 9}, 1255--1261.

\bibitem[\protect\BCAY{Tsai\ \BBA\ Kwee}{Tsai\ \BBA\
  Kwee}{2011}]{Tsai_weighting_2011}
Tsai, F.~S.\BBACOMMA\  \BBA\ Kwee, A.~T. \BBOP2011\BBCP.
\newblock \BBOQ Experiments in term weighting for novelty mining\BBCQ\
\newblock {\Bem Expert Systems with Applications}, {\Bem 38\/}(11), 14094 --
  14101.

\bibitem[\protect\BCAY{Tsai, Tang,\ \BBA\ Chan}{Tsai et~al.}{2010}]{sl4}
Tsai, F.~S., Tang, W., \BBA\ Chan, K.~L. \BBOP2010\BBCP.
\newblock \BBOQ Evaluation of novelty metrics for sentence-level novelty
  mining\BBCQ\
\newblock {\Bem Inf. Sci.}, {\Bem 180\/}(12), 2359--2374.

\bibitem[\protect\BCAY{Verheij, Kleijn, Frasincar,\ \BBA\ Hogenboom}{Verheij
  et~al.}{2012}]{Verheij_2012}
Verheij, A., Kleijn, A., Frasincar, F., \BBA\ Hogenboom, F. \BBOP2012\BBCP.
\newblock \BBOQ A comparison study for novelty control mechanisms applied to
  web news stories\BBCQ\
\newblock In Zhong, N., Gong, Z., ming Cheung, Y., Lingras, P., Szczepaniak,
  P.~S., \BBA\ Suzuki, E.\BEDS, {\Bem The 2012 IEEE/WIC/ACM International
  Conference on Web Intelligence (WI 2012)}, \BPGS\ 431--436. {IEEE Computer
  Society}.

\bibitem[\protect\BCAY{Yang, Zhang, Carbonell,\ \BBA\ Jin}{Yang
  et~al.}{2002}]{topic}
Yang, Y., Zhang, J., Carbonell, J., \BBA\ Jin, C. \BBOP2002\BBCP.
\newblock \BBOQ Topic-conditioned novelty detection\BBCQ\
\newblock In {\Bem 8th ACM SIGKDD international conference on Knowledge
  discovery and data mining}, KDD '02, \BPGS\ 688--693. ACM.

\bibitem[\protect\BCAY{Zhang, Zi,\ \BBA\ Wu}{Zhang
  et~al.}{2007}]{Zhang_NED_2007}
Zhang, K., Zi, J., \BBA\ Wu, L.~G. \BBOP2007\BBCP.
\newblock \BBOQ New event detection based on indexing-tree and named
  entity\BBCQ\
\newblock In {\Bem 30th annual international ACM SIGIR conference on Research
  and development in information retrieval}, SIGIR '07, \BPGS\ 215--222, New
  York, NY, USA. ACM.

\bibitem[\protect\BCAY{Zhang, Callan,\ \BBA\ Minka}{Zhang
  et~al.}{2002}]{filtering}
Zhang, Y., Callan, J., \BBA\ Minka, T. \BBOP2002\BBCP.
\newblock \BBOQ Novelty and redundancy detection in adaptive filtering\BBCQ\
\newblock In {\Bem 25th annual international ACM SIGIR conference on Research
  and development in information retrieval}, SIGIR '02, \BPGS\ 81--88. ACM.

\bibitem[\protect\BCAY{Zhang\ \BBA\ Tsai}{Zhang\ \BBA\ Tsai}{2009}]{sl3}
Zhang, Y.\BBACOMMA\  \BBA\ Tsai, F.~S. \BBOP2009\BBCP.
\newblock \BBOQ Combining named entities and tags for novel sentence
  detection\BBCQ\
\newblock In {\Bem WSDM '09 Workshop on Exploiting Semantic Annotations in
  Information Retrieval}, ESAIR '09, \BPGS\ 30--34. ACM.

\end{thebibliography}
